\documentclass[11pt,a4paper]{article}
\usepackage{amsfonts}
\usepackage{amsmath,amssymb}
\usepackage{epsfig,graphicx}

\input{tcilatex}

\begin{document}

\begin{center}
\bigskip

\bigskip {\huge Standard Model with Partial Gauge Invariance}{\LARGE \ }

\bigskip

\bigskip

\bigskip

{\LARGE \ }

{\LARGE \ }

\textbf{J.L.~Chkareuli}$^{1,2}$ and \textbf{\ Z. Kepuladze}$^{1,2}$\textbf{\ 
}\bigskip

$^{1}$\textit{Center for Elementary Particle Physics, ITP,} \textit{Ilia
State University, 0162 Tbilisi, Georgia}

$^{2}$\textit{Andronikashvili} \textit{Institute of Physics, 0177 Tbilisi,
Georgia\ \ \ }

\bigskip \bigskip \bigskip \bigskip \bigskip \bigskip \bigskip \bigskip
\bigskip

\textbf{Abstract}

\bigskip
\end{center}

\bigskip

We argue that an exact gauge invariance may disable some generic features of
the Standard Model which could otherwise manifest themselves at high
energies. One of them might be related to the spontaneous Lorentz invariance
violation (SLIV) which could provide an alternative dynamical approach to
QED and Yang-Mills theories with photon and non-Abelian gauge fields
appearing as massless Nambu-Goldstone bosons. To see some key features of
the new physics expected we propose partial rather than exact gauge
invariance in an extended SM framework. This principle applied, in some
minimal form, to the weak\ hypercharge gauge field $B_{\mu }$ and its
interactions leads to SLIV with $B$ field components appearing as the
massless Nambu-Goldstone modes, and provides a number of distinctive Lorentz
beaking effects. Being naturally \ suppressed at low energies they may
become detectable in high energy physics and astrophysics. Some of the most
interesting SLIV\ processes are considered in significant detail.

\thispagestyle{empty}\newpage

\section{Introduction}

It is now generally accepted that internal gauge symmetries form the basis
of modern particle physics being most successfully realized within the
celebrated Standard Model of quarks and leptons and their fundamental
strong, weak and electromagnetic interactions.

On the other hand, postulated local gauge symmetries, unlike global
symmetries, represent redundancies of the description of a theory rather
than being \textquotedblleft true\textquotedblright\ symmetries. Indeed, as
has been discussed time and again (for some instructive example, see \cite%
{11}), the very existence of an exact gauge invariance means that there are
more field variables in the theory than are physically necessary. Usually,
these supefluous degrees of freedom are eliminated by some gauge-fixing
conditions which have no physical meaning in themselves and actually are put
by hand. Instead, one could think that these extra variables would vary
arbitrarily with the time so that they could be made to serve in description
of some new physics.

One of possible ways for such new physics to appear may be linked to the
idea that local symmetries and the associated masslessness of gauge bosons
have in essence a dynamical origin rather than being due to a fundamental
principle, as was widely contemplated over the last several decades \cite%
{4,5,61}. By analogy with a dynamical origin of massless scalar particle
excitations, which is very well understood in terms of spontaneously broken
global internal symmetries, the origin of massless gauge fields as vector
Nambu-Goldstone (NG) bosons could be related to the spontaneous violation of
Lorentz invariance which is in fact the minimal spacetime global symmetry
underlying the elementary particle physics. This approach providing a
valuable alternative framework to quantum electrodynamics and Yang-Mills
theories has gained new impetus \cite{61} in recent years\footnote{%
Independently of the problem of the origin of local symmetries, Lorentz
violation in itself has attracted considerable attention as an interesting
phenomenological possibility that may be probed in direct Lorentz
non-invariant, while gauge invariant, extensions of QED and Standard Model
(SM) \cite{refs,2,3}.}.

However, in contrast to the spontaneous internal symmetry violation which is
readily formulated in gauge invariant theories, the spontaneous Lorentz
invariance violation (SLIV) implies in general an explicit breakdown of
gauge invariance in order to physically manifest itself. Indeed, the
simplest model for SLIV is given by a conventional QED type Lagrangian
extended by an arbitrary vector field potential energy terms 
\begin{equation}
U(A)=\frac{\lambda }{4}\left( A_{\mu }A^{\mu }-n^{2}M^{2}\right) ^{2}
\label{pol}
\end{equation}%
which are obviously forbidden by a strict $U(1)$ gauge invariance of the
starting Lagrangian. Here $n_{\mu }$ ($\mu =0,1,2,3$) is a properly-oriented
unit Lorentz vector, $n^{2}=n_{\mu }n^{\mu }=\pm 1$, while $\lambda $ and $%
M^{2}$ are, respectively, dimensionless and mass-squared dimensional
positive parameters. This potential means that the vector field $A_{\mu }$
develops a constant background value $\langle A_{\mu }\rangle =n_{\mu }M$\
and Lorentz symmetry $SO(1,3)$ breaks at the scale $M$ down to $SO(3)$ or $%
SO(1,2)$ depending on whether $n_{\mu }$ is time-like ($n_{\mu }^{2}>0$) or
space-like ($n_{\mu }^{2}<0$). Expanding the vector field around this vacuum
configuration, 
\begin{equation}
A_{\mu }(x)=n_{\mu }(M+\phi )+a_{\mu }(x)~,\text{ \ }n_{\mu }a^{\mu }=0
\label{2}
\end{equation}%
one finds that the $a_{\mu }$ field components, which are orthogonal to the
Lorentz violating direction $n_{\mu }$, describe a massless vector
Nambu-Goldstone boson, while the $\phi (x)$ field corresponds to a Higgs
mode. This minimal polynomial extension of QED, being sometimes referred to
as the \textquotedblleft bumblebee\textquotedblright\ model, is\ in fact the
prototype SLIV model intensively discussed in the literature (see \cite%
{bluhm} and references therein).

So, if one allows the vector field potential energy like terms (\ref{pol})
to be included into the properly modified QED Lagrangian, the time-like or
space-like SLIV could unavoidably hold thus leading to photon as the
massless NG boson in the symmetry broken SLIV phase. If this SLIV pattern is
taken as some generic feature of QED the gauge principle should be properly
weakened, otherwise this feature might be disabled. It is clear that this
type of reasoning can be equally applied to any model possesing, among
others, some local $U(1)$ symmetry which is properly broken by the
corresponding gauge field terms in the Lagrangian. Remarkably, just the $%
U(1) $ local symmetry case with its gauge field (\ref{2}) possessing one
Higgs and three Goldstone components (being equal to number of broken
Lorentz generators) appears to be optimally fitted for the physically
valuable SLIV mechanism\footnote{%
Note in this connection that SLIV through the condensation of non-Abelian
vector fields would lead to a spontaneous breakdown of internal symmetry as
well (see some discussion in \cite{jej}) that could make our consideration
much more complicated.}. Actually, if things were arranged in this way, one
could have indeed an extremely attractive dynamical alternative to
conventional QED and/or Standard Model that could be considered in itself as
some serious motivation for a status of an overall gauge symmetry in them to
be properly revised.

In this connection, \textit{we propose partial rather than exact gauge\
invariance in the Standard Model according to which, while the electroweak
theory is basically }$SU(2)\times U(1)_{Y}$\textit{\ gauge invariant being
constructed from ordinary covariant derivatives of all fields involved, the }%
$U(1)_{Y}$\textit{\ hypercharge gauge field }$B_{\mu }$\textit{\ field is
allowed to form all possible polynomial \textit{\ couplings on its own and
with other fields invariants}. }So\textit{, }the new terms in the SM
Lagrangian, conditioned by the partial gauge invariance, may generally have
a form%
\begin{equation}
-U(B)+B_{\mu }\Upsilon ^{\mu }(f,h,g)+B_{\mu }B_{\nu }\Theta ^{\mu \nu
}(f,h,g)+\text{ }\cdot \cdot \cdot \   \label{ntt}
\end{equation}%
where $U(B)$ contains all possible $B$ field potential energy terms, the
second term in (\ref{ntt}) consists of all vector type couplings with the SM
fields involved (including left-handed and right-handed fermions $f$, Higgs
field $h$ and gauge fields $g$), the third term concerns possible tensor
like couplings, and so on. These new terms (with all kinds of the $%
SU(3)_{c}\times SU(2)\times U(1)_{Y}$ gauge invariant tensors $\Upsilon
^{\mu },$ $\Theta ^{\mu \nu }$ etc.) "feel" only $B$ field gauge
transformations while remaining invariant under gauge transformations of all
other fields. Ultimately, just their sensitivity to the $B$ field gauge
transformations leads to physical Lorentz violation in SM. Indeed, the
constant part of the vector field SLIV pattern (\ref{2}) can be treated in
itself as some gauge transformation with gauge function linear in
coordinates, $\omega (x)=$ $(n_{\mu }x^{\mu })M$, and therefore, this
violation may physically emerge only through the terms like those in (\ref%
{ntt}) which only possess partial gauge invariance (PGI).

The paper is organized as follows. In section 2 we make some natural
simlification of a general PGI conjecture given above in (\ref{ntt}) and
find an appropriate minimal form for PGI. We exclude the accompanying SLIV
Higgs component in the theory going to the nonlinear realization of Lorentz
symmetry and propose that all possible vector and tensor couplings in the
PGI expansion (\ref{ntt}) are solely determined by the SM Noether currents.
When only vector couplings are taken in (\ref{ntt}) this leads to the simple
nonlinear Standard Model (NSM) which is considered in detail in the next
section 3, and its physical Lorentz invariance and observational equivalence
to the conventional SM is explicitly demonstrated. In section 4 we will
mainly be focused on the extended NSM (ENSM) with the higher dimensional
tensor coupling terms included. By contrast, they lead to the physical SLIV
with a variety distinctive Lorentz breaking effects in a laboratory some of
which are considered in detail. And, finally, in section 5 we conclude.

\section{Partial Gauge Invariance Simplified}

Generally, the PGI conjecture, as formulated in (\ref{ntt}), may admit too
many extra terms in the SM Lagrangian. However, one can have somewhat more
practical choice for PGI. This is related to the way SLIV is realized in SM
and a special role which two SM Noether currents, namely, the total
hypercharge current and the total energy-momentum tensor of all fields
involved\footnote{%
Notice that these currents are proposed to be used in the form they have in
an ordinary rather than extended Standard Model (for further discussion, see
subsection 2.3).}, may play in formulation of the PGI conjecture (\ref{ntt}%
). We give below a brief discussion of each of terms in (\ref{ntt}) and try
to make some possible simplifications.

\subsection{Nonlinear Lorentz realization}

\ The first thing of interest in (\ref{ntt}) is the potential energy terms $%
U(B)$ for the SM\ hypercharge gauge field $B_{\mu }$, which are like those
we had in (\ref{pol}) for QED. Though generally just these terms cause a
spontaneous Lorentz violation, their physical effects are turned out to be
practically insignificant unless one considers some special SLIV interplay
with gravity \cite{bluhm, grip} at the super-small distances, or a possible
generation of the SLIV topological defects in the very early universe \cite%
{ckv}. Actually, as in the pure SLIV\ QED case \cite{kraus}, one has an
ordinary Lorentz invariant low energy physics in an effective SLIV\ SM
theory framework. The only Lorentz breaking effects may arise from radiative
corrections due to the essentially decoupled superheavy (with the SLIV scale
order mass) Higgs component contributions, which are generally expected to
be negligibly small at lower energies.

For more clearness and simplicity, we completely exclude this vector field
Higgs component in the theory going to the nonlinear realization of Lorentz
symmetry through the nonlinear $\sigma $-model for the hypercharge gauge
field $B_{\mu }$, just as it takes place in the original nonlinear $\sigma $%
-model \cite{8} for pions\footnote{%
This correspondence with the nonlinear $\sigma $ model for pions may be
somewhat suggestive, in view of the fact that pions are the only presently
known NG bosons and their theory, chiral dynamics \cite{8}, is given by the
nonlinearly realized chiral $SU(2)\times SU(2)$ symmetry rather than by an
ordinary linear $\sigma $ model.}. Actually, for the pure QED case this has
been done by Nambu long ago \cite{6} (see also \cite{7} for some recent
discussion). Doing so in the SM framework, particularly in its hypercharge
sector, one immediately comes to the $B$ field constraint\footnote{%
Actually, as in the pion model, one can go from the linear model for SLIV to
the nonlinear one taking the corresponding potential, similar to the
potential (\ref{pol}), to the limit $\lambda \rightarrow \infty $.} 
\begin{equation}
B_{\mu }^{2}=n^{2}M^{2}\text{ .}  \label{const}
\end{equation}%
This constraint provides in fact the genuine Goldstonic nature of the
hypercharge gauge field appearing at the SLIV scale $M$, as could easily be
seen from an appropriate $B$ field parametrization, 
\begin{equation}
B_{\mu }=b_{\mu }+\frac{n_{\mu }}{n^{2}}(M^{2}-n^{2}b_{\nu }^{2})^{\frac{1}{2%
}},\text{ \ }n_{\mu }b^{\mu }=0\text{\ \ }  \label{par}
\end{equation}%
with the pure NG modes $b_{\mu }$ and an effective Higgs mode (or the $B$
field component in the vacuum direction) being given by the square root in (%
\ref{par}). Indeed, both of these SLIV patterns in the SM framework, linear
and nonlinear, are equivalent in the infrared energy domain, where the Higgs
mode is considered to be infinitely massive. We consider for what follows
just the nonlinear SM (or NSM, as we call it hereafter) where SLIV is
related to an explicit nonlinear constraint put on the hypercharge gauge
field (\ref{const}) rather than to a presence of its potential energy terms
in the SM Lagrangian. We show later in section 3 that this theory with the
corresponding Lagrangian $\mathcal{L}_{NSM}$ written in the pure NG modes $%
b_{\mu }$ (\ref{par}) is physically equivalent to an ordinary SM theory.

\subsection{The minimal PGI}

Further, we propose for the second term in the PGI extension of SM (\ref{ntt}%
) that it is solely given by $B$ field dimensionless couplings with the
total hypercharge current $J^{\mu }(f,h,g)$\ of all matter fields involved.
Namely, $\Upsilon ^{\mu }$ in (\ref{ntt}) is replaced by $\mathfrak{g}%
^{\prime }J^{\mu }$ , where $\mathfrak{g}^{\prime }$ is some coupling
constant. However, the inclusion of these couplings into the NSM Lagrangian $%
\mathcal{L}_{NSM}$ would only redefine the original hypercharge gauge
coupling constant $g^{\prime }$ which is in essence a free parameter in SM.
This means that for the basic theory with dimensionless coupling constants
the partial gauge invariance is really indistinguishable from an ordinary
gauge invariance due to which SLIV can be gauged away in the basic NSM, as
we shall see in the next section. Otherwise, the large Lorentz breaking
effects would make the whole model absolutely irrelevant. From the above
reasoning the second term in (\ref{ntt}) will be simply omitted in the
subsequent discussion.

Meanwhile, a clear signal of the physical Lorentz violation inevitably
occurs when one goes beyond the minimal theory to also activate the\ higher
dimensional tensor couplings in (\ref{ntt}) that leads to the extended
nonlinear SM (or ENSM). For further simplicity, these couplings are proposed
to be determined solely by the total energy-momentum tensor $T^{\mu \nu }$
of all fields involved, namely, $\Theta ^{\mu \nu }=$ $(\alpha
/M_{P}^{2})T^{\mu \nu }$. So, the lowest order ENSM which conforms with the
chiral nature of SM and all accompanying global and discrete symmetries, is
turned out to include the dimension-6 couplings of the \ type

\begin{equation}
\mathcal{L}_{ENSM}=\mathcal{L}_{NSM}+\alpha \dfrac{B_{\mu }B_{\nu }}{%
M_{P}^{2}}T^{\mu \nu }(f,g,h)\text{ }  \label{Ltot}
\end{equation}%
describing at the Planck scale $M_{P}$ the extra interactions of the
hypercharge\ gauge fields with the energy-momentum tensor bilinears of
matter fermions, and gauge and Higgs bosons (with the dimensionless coupling
constant $\alpha $ indicated), respectively. The $T^{\mu \nu }$ tensor in (%
\ref{Ltot}) is proposed to be symmetrical (in spacetime indices) and $%
SU(3)_{c}\times SU(2)\times U(1)_{Y}$ gauge invariant according to our basic
conjecture (\ref{ntt}). So, the physical Lorentz violation, in a form that
follows from this minimal PGI determined by the SM Noether currents$^{3}$,
appears to be naturally suppressed thus being in a reasonable compliance
with current experimental bounds. Nonetheless, as we show in section 4, the
extra couplings in (\ref{Ltot}) may lead, basically through the deformed
dispersion relations of all matter and gauge fields involved, to a new class
of processes which could still be of a distinctive observational interest in
high energy physics and astrophysics. As to the higher dimensional couplings
in (\ref{ntt}), we assume that they are properly suppressed or even
forbidden if one takes a minimal choice for PGI (\ref{Ltot}) to which we
follow here.

\subsection{How the minimal PGI works}

Now, one can readily see that the SM hypercharge current $J^{\mu }$\ and
energy-momentum tensor $T^{\mu \nu }$ used above as the only buildding
blocks for the simplified version of ENSM\ (\ref{Ltot}) may really determine
some minimal gauge symmetry breaking mechanism in the theory. Indeed, they
are changed when SM is modified by the tensor type couplings so that one has
new conserved Noether currents $J^{\prime \mu }$ and $T^{\prime \mu \nu }$
in ENSM 
\begin{eqnarray}
J^{\prime \mu } &=&J^{\mu }+\varkappa B^{\mu }B_{\rho }J^{\rho },\text{ \ \ }
\label{cur} \\
T^{\prime \mu \nu } &=&T^{\mu \nu }+\frac{\varkappa }{2}(B^{\mu }B_{\rho
}T^{\rho \nu }+B^{\nu }B_{\rho }T^{\rho \mu })  \notag
\end{eqnarray}%
(where $\varkappa $ stands for $\alpha /M_{P}^{2}$), while the old currents $%
J^{\mu }$ and $T^{\mu \nu }$ participating in the couplings (\ref{Ltot}) are
only approximately conserved, $\partial _{\mu }J^{\mu }=O(\varkappa )$\ and $%
\partial _{\mu }T^{\mu \nu }=O(\varkappa )$. Nonetheless, this appears
enough to have, in turn, the "almost" gauge invariant field equations in
ENSM. Indeed, the Lagrangian density (\ref{Ltot}) varies to the taken
accuracy into some total derivative%
\begin{equation}
\delta \mathcal{L}_{ENSM}=\varkappa \partial _{\mu }(R_{\nu }T^{\mu \nu
})+O(\varkappa ^{2})  \label{apr}
\end{equation}%
where $R_{\nu }$ stands for the properly defined integral function%
\begin{equation}
R_{\nu }(x)=\int^{x}dx^{\rho }(B_{\rho }\partial _{\nu }\omega +B_{\nu
}\partial _{\rho }\omega +\partial _{\rho }\omega \partial _{\nu }\omega )
\label{R}
\end{equation}%
conditioned by the corresponding $B$ field gauge transformations%
\begin{equation}
B_{\mu }\rightarrow B_{\mu }+\partial _{\mu }\omega (x)\text{ .}  \label{tr}
\end{equation}%
The gauge function $\omega (x)$ is an arbitrary function, only being
restricted by the requirement to conform with the $B$ field constraint (\ref%
{const}) 
\begin{equation}
(B_{\mu }+\partial _{\mu }\omega )(B^{\mu }+\partial ^{\mu }\omega
)=n^{2}M^{2}\text{ ,}  \label{hj}
\end{equation}%
due to which the introduced integral function $R_{\nu }$ (\ref{R}) has to be
divergenceless%
\begin{equation}
\partial ^{\nu }R_{\nu }=B_{\mu }^{2}-n^{2}M^{2}=0\text{ .}
\end{equation}%
Should a solution to the constraint equation (\ref{hj}) exist for some class
of finite gauge functions $\omega (x)$ the reverse would also be true:
requiring the above approximate gauge invariance (\ref{apr}) of the
Lagrangian under transformations (\ref{tr}) one comes to the minimal ENSM (%
\ref{Ltot}). The actual physical equivalence of NSM determined by the SLIV
constraint (\ref{const}) to an ordinary SM theory, that is explicitly
demonstrated in section 3, shows that such gauge function may really exist%
\footnote{%
This confirms that the constraint (\ref{const}) in itself may well only be
some particular gauge choice in SM to which just NSM corresponds (for more
discussion, see section 3).}. So, the partial gauge invariance in a minimal
form taken above tends to appear reasonably well defined at least at the
classical level.

\subsection{The metric expansion viewpoint}

It is conceivable, on the other hand, that the extra interaction terms in $%
\mathcal{L}_{ENSM}$\ (\ref{Ltot}) might arise as remnants of some operator
expansion of the metric tensor $g_{\mu \nu }(x)$ into all possible
tensor-valued covariants which could generally appear in quantum gravity.
For metric correlated with the total energy-momentum tensor $T^{\mu \nu
}(f,g,h)$\ of SM (or NSM\ in the nonlinear Lorentz realization case) this
expansion 
\begin{equation}
g_{\mu \nu }=\eta _{\mu \nu }+\mathfrak{h}_{\mu \nu }/M_{P}+\alpha B_{\mu
}B_{\nu }/M_{P}^{2}+\beta \mathbf{W}_{\mu }\mathbf{W}_{\nu }/M_{P}^{2}+\cdot
\cdot \cdot  \label{g}
\end{equation}%
may include, along with the Minkowski metric tensor $\eta _{\mu \nu }$ and
graviton field $\mathfrak{h}_{\mu \nu }$, as is usually taken in a weak
gravity approximation, the SM gauge and matter field covariants as well ($%
\alpha ,$ $\beta ,...$ are coupling constants). As a result, once SLIV
occurs with $B$ field developing a constant background value (\ref{par})\
the conventional SM interactions appear to be significantly modified at
small distances presumably controlled by quantum gravity.

\subsection{Running to low energies}

Now let us concretize the form of the minimal ENSM theory given above (\ref%
{Ltot}). First of all, note that the Lagrangian containing part ($-\eta
^{\mu \nu }\mathcal{L}_{NSM}$) in the total energy-momentum tensor $T^{\mu
\nu }(f,g,h)$ appears unessential since it only leads to a proper
redefinition of all fields involved regardless their properties under SM.
Actually, the contraction of \ this part with the shifted hypercharge gauge
field $B_{\mu }$ in (\ref{par}) gives in the lowest order the universal
factor 
\begin{equation}
1-\alpha \dfrac{M^{2}n^{2}}{M_{P}^{2}}
\end{equation}%
to the whole nonlinear SM Lagrangian $\mathcal{L}_{NSM}$ considered. So, we
will consider only "the Lagrangian subtracted" energy-momentum tensor $%
T^{\mu \nu }$ in what follows (leaving the former notation for it).

The more significant point concerns the running of the coupling constant $%
\alpha $ for the basic extra interaction of ENSM (\ref{Ltot}). It is clear
that even if one starts with one universal constant at the Planck scale $%
M_{P}$, it will appear rather different for matter fermions ($\alpha _{f}$),
gauge fields ($\alpha _{g}$) and Higgs boson ($\alpha _{h}$) being
appropriately renormalized when running down to lower energies. Moreover,
each of these constants is further split for different fermion and gauge
multiplets in SM that is determined, in turn, by the corresponding radiative
corrections. For example, one could admit that quarks and leptons have equal 
$\alpha $-coupling ($\alpha _{f}$) in the Planck scale limit. However, due
to radiative corrections this coupling constant may split into two ones -
one for quarks ($\alpha _{q}$) and another for leptons ($\alpha _{l}$),
respectively, that could be in principle calculated. Apart from that, there
appear two more coupling constants, namely, those for left-handed quarks and
leptons\ ($\alpha _{q_{L}},\alpha _{l_{L}}$) and right-handed ones ($\alpha
_{q_{R}},\alpha _{l_{R}}$). We will take into account some difference
between $\alpha $-couplings of quarks and leptons but will ignore such a
difference for left-handed and right-handed fermions of the same species.
Indeed, the associated radiative corrections, which basically appear due to
the chirality-dependent weak interactions in SM, are expected to be
relatively small. So, practically there are only four effective coupling
constants at normal laboratory energies, $\alpha _{f}$ ($f=q,l$), $\alpha
_{g}$ and $\alpha _{h}$, in the theory with one quark-lepton family. In
other words, the total energy-momentum tensor $T^{\mu \nu }(f,g,h)$ in the
basic ENSM coupling (\ref{Ltot}) breaks into the sum 
\begin{equation}
T^{\mu \nu }(f,g,h)=\frac{\alpha _{f}}{\alpha }T_{f}^{\mu \nu }+\frac{\alpha
_{g}}{\alpha }T_{g}^{\mu \nu }+\frac{\alpha _{h}}{\alpha }T_{h}^{\mu \nu }
\label{fgh}
\end{equation}%
of the energy-momentum tensors of matter fermions, and gauge and Higgs
bosons, respectively, when going from the Planck scale down to low energies.
However, the different fermion quark-lepton families may still have rather
different $\alpha $-couplings that could eventually lead to the
flavor-changing processes in our model (some interesting examples are
discussed in section 4)\footnote{%
Note that, apart from the proposed generic Planck scale unification of the
PGI couplings in ENSM (\ref{Ltot}) there could be some intermediate grand
unification \cite{moh} and/or family unification (for some example, see \cite%
{su8}) in the theory. These extra symmetries will also influence their
running down to low energies.}.

\section{Nonlinear Standard Model}

In contrast to the spontaneous violation of internal symmetries, SLIV seems
not to necessarily imply a physical breakdown of Lorentz invariance. Rather,
when appearing in a minimal gauge theory framework, this may eventually
result in a noncovariant gauge choice in an otherwise gauge invariant and
Lorentz invariant theory. This is what just happens in a simple class of QED
type models \cite{6,7} having from the outset a gauge invariant form, in
which SLIV is realized through the "length-fixing" field constraint of the
type (\ref{const}) rather than due to some vector field potential energy
terms. Remarkably, this type of model makes the vector Goldstone boson a
true gauge boson (photon), whereas the physical Lorentz invariance is left
intact. Indeed, despite an evident similarity with the nonlinear $\sigma $%
-model for pions, the nonlinear QED theory ensures that all the physical
Lorentz violating effects prove to be non-observable. Particularly, it was
shown, first only in the tree approximation \cite{6}, that the nonlinear
constraint (\ref{const}) implemented as a supplementary condition into the
standard QED Lagrangian appears in fact as a possible gauge choice for the
vector field $A_{\mu }$. At the same time the $S$-matrix remains unaltered
under such a gauge convention. Really, this nonlinear QED contains a
plethora of Lorentz and $CPT$ violating couplings when it is expressed in
terms of the pure NG photon modes according to the constraint condition
being similar to (\ref{const}). However, the contributions of these Lorentz
violating couplings to physical processes completely cancel out among
themselves. So, SLIV was shown to be superficial as it affects only the
gauge of the vector potential $A_{\mu }$, at least in the tree approximation 
\cite{6}.

Some time ago, this result was extended to the one-loop approximation \cite%
{7}. It was shown that the constraint like (\ref{const}), having been
treated as a nonlinear gauge choice for the $A_{\mu }$ field at the tree
(classical) level, remains as a gauge condition when quantum effects in
terms of the loop diagrams are taken into account as well. So, one can
conclude that physical Lorentz invariance is left intact in the one-loop
approximation in the nonlinear QED taken in the flat Minkowski spacetime.

We consider here in this section the nonlinear Standard Model (or NSM)
treated merely as SM with the nonlinear constraint (\ref{const}) put on the
hypercharge gauge field $B_{\mu }$. We show that NSM despite many generic
complications involved (like as the spontaneos breaking of the internal $%
SU(2)\times U(1)_{Y}$ symmetry, the diverse particle spectrum, mixings in
gauge and matter sectors, extension by the PGI vector couplings\footnote{%
Note that the extension of NSM by the extra vector couplings in (\ref{ntt})
which according to the minimal PGI (section 2) are solely determined by the
total hypercharge current are simply absorbed in NSM only leading to the
redifinition of the hypercharge coupling constant.} etc.) appears
observationally equivalent to the ordinary SM, just like what happens in the
above mentioned nonlinear QED case. Actually, due to the SLIV\ constraint (%
\ref{const}), physical $B$ field components convert into the massless NG
modes which, after an ordinary electroweak symmetry breaking, mix with a
neutral $W^{3}$ boson of $SU(2)$ leading, as usual, to the massless photon
and massive $Z$ boson. When expressed in terms of the pure NG modes NSM,
like the nonlinear QED, contains a variety of Lorentz and $CPT$ violating
couplings. Nonetheless, all SLIV effects turn out to be strictly cancelled
in all lowest order processes some of which are considered in detail below.

\subsection{Hypercharge vector Goldstone boson}

We start with the Standard Model where, for simplicity, we restrict
ourselves to the electron family only%
\begin{equation}
L=\left( 
\begin{array}{c}
\nu _{e} \\ 
e%
\end{array}%
\right) _{L},\ e_{R}
\end{equation}%
that can be then straightforwardly extended to all matter fermions\
observed. For the starting hypercharge gauge field $B_{\mu }$ expessed in
terms of its Goldstone counterpart $b_{\mu }$ (\ref{par}) one has  in the
leading order in the inverse SLIV scale $1/M$  $\ \ \ \ \ \ \ \ $%
\begin{equation}
B_{\mu }=b_{\mu }+\dfrac{n_{\mu }}{n^{2}}M-\dfrac{b_{\nu }^{2}}{2M}n_{\mu }%
\text{ }  \label{BB}
\end{equation}%
so that in the same order the hypercharge field stress-tensor\ $B_{\mu \nu }$
amounts to

\begin{equation}
B_{\mu \nu }=\partial _{\mu }B_{\nu }-\partial _{\nu }B_{\mu }=b_{\mu \nu }-%
\frac{1}{2M}\left( n_{\nu }\partial _{\mu }-n_{\mu }\partial _{\nu }\right)
\left( b_{\rho }\right) ^{2}\text{ }.  \label{B}
\end{equation}%
When one also introduce the NG modes $b_{\mu }$ in the hypercharge covariant
derivatives for all matter fields involved one eventually comes to the
essentially nonlinear\ $b$ field theory sector in the Standard Model due to
which it is now called the nonlinear SM or NSM.

This model might seem unacceptable since it contains, among other terms, the
inappropriately large (the SLIV scale $M$ order) Lorentz violating fermion
and Higgs fields bilinears which appear when the starting $B$ field
expansion (\ref{BB}) is applied to the corresponding couplings in SM.
However, due to partial gauge invariance, according to which all matter
fields remain to possess the covariant derivatives, these bilinears can be
gauged away by making an appropriate field redefinition according to 
\begin{equation}
(L,e_{R},H)\longrightarrow (L,e_{R},H)\exp (i\frac{Y_{L,R,H}}{2}g^{\prime
}n^{2}M(n_{\mu }x^{\mu })  \label{red}
\end{equation}%
So, one  eventually comes to the nonlinear SM\ Lagrangian   
\begin{equation}
\mathcal{L}_{NSM}=\mathcal{L}_{SM}(B_{\mu }\rightarrow b_{\mu })+\mathcal{L}%
_{nSM}  \label{SM}
\end{equation}%
where the conventional SM part being expressed in terms of the the
hypercharge NG vector boson $b_{\mu }$ is presented in $\mathcal{L}%
_{SM}(B_{\mu }\rightarrow b_{\mu })$, while its essentially nonlinear
couplings are collected in $\mathcal{L}_{nSM}$ written in the taken order $%
O(1/M)$ as  
\begin{eqnarray}
2M\mathcal{L}_{nSM} &=&-(n\partial )b_{\mu }\partial ^{\mu }(b_{\nu }^{2})+%
\dfrac{1}{2}g^{\prime }b_{\nu }^{2}\overline{L}\gamma ^{\mu }n_{\mu
}L+g^{\prime }b_{\nu }^{2}\overline{e}_{R}\gamma ^{\mu }n_{\mu }e_{R}
\label{lint} \\
&&-\frac{i}{2}g^{\prime }b_{\nu }^{2}\left[ H^{+}(n_{\mu }\partial ^{\mu
}H)-(n_{\mu }\partial ^{\mu }H^{+})H\right]   \notag
\end{eqnarray}%
Note that the SLIV conditioned "gauge" $n_{\mu }b^{\mu }=0$ (\ref{par}) for
the $b$-field is imposed everywhere in the Lagrangian $\mathcal{L}_{NSM}$.
Moreover, we take the similar axial gauge for $W^{i}$ bosons of $SU(2)$ so
as to have together%
\begin{equation}
n_{\mu }W^{i\mu }=0\text{ , \ }n_{\mu }b^{\mu }=0\text{\ \ .}  \label{ag}
\end{equation}%
in what follows. As a result, all terms containing contraction of the unit
vector $n_{\mu }$ with electroweak boson fields will vanish in the $\mathcal{%
L}_{NSM}$.

We see later that NSM, despite the presence of particular Lorentz and $CPT$
violating couplings in its essentially nonlinear part (\ref{lint}), does not
lead in itself to the physical Lorentz violation until the extra PGI
couplings appearing in the ENSM Lagrangian (\ref{Ltot}) start working.

\subsection{Electroweak symmetry breaking in NSM}

At much lower energies than the SLIV scale $M$ a conventional spontaneous\
breaking of the internal symmetry $SU(2)\times U(1)_{Y}$ naturally holds in\
NSM. This appears when the Higgs field $H$ acquires the constant background
value through its potential energy terms 
\begin{equation}
U(H)=\mu _{H}^{2}H^{+}H+(\lambda /2)(H^{+}H)^{2}\text{ , \ }\mu _{H}^{2}<0
\label{pe}
\end{equation}%
in the electroweak Lagrangian. Due to the overall axial gauge adopted (\ref%
{ag}) there is no more a gauge freedom\footnote{%
This kind of SM with all gauge bosons taken in the axial gauge was earlier
studied \cite{dams} in an ordinary Lorentz invariant framework. Also, the
SLIV condioned axially gauged vector fields in the spontaneously broken
massive QED\ was considered in \cite{7}.} in NSM to exclude extra components
in the $H$ doublet. So, one can parametrize it in the following general form

\begin{equation}
H\equiv \frac{1}{\sqrt{2}}\left( 
\begin{array}{c}
\phi  \\ 
(h+V)e^{i\xi /V}%
\end{array}%
\right) \text{ , \ \ \ }V=(-\mu _{H}^{2}/\lambda )^{1/2}  \label{h}
\end{equation}%
The would-be scalar Goldstone bosons, given by the real $\xi $ and complex $%
\phi (\phi ^{\ast })$ fields$,$ mix generally with \ $Z$ \ boson and\ $%
W(W^{\ast })$ boson components, respectively. To see these mixings one has
to write all bilinear terms stemming from the starting Higgs doublet
Lagrangian which consists of its covariantized kinetic term $\left\vert
D_{\mu }H\right\vert ^{2}$ and the potential energy part (\ref{pe}). Putting
them all together one comes to

\begin{equation}
(\partial ^{\mu }h)^{2}/2+\mu _{h}^{2}h^{2}/2+\left\vert M_{W}W_{\mu
}-i\partial _{\mu }\phi \right\vert ^{2}+(M_{Z}Z_{\mu }+\partial _{\mu }\xi
)^{2}/2  \label{ssww}
\end{equation}%
where we have used the usual expression for Higgs boson mass $\mu
_{h}^{2}=\lambda \left\vert \mu _{H}^{2}\right\vert $, and also the
conventional expressions for $W$ and $Z$\ bosons 
\begin{equation}
(W_{\mu },W_{\mu }^{\ast })=(W_{\mu }^{1}\pm iW_{\mu }^{2})/\sqrt{2}\text{ ,
\ \ }Z_{\mu }=\cos \theta ~W_{\mu }^{3}-\sin \theta ~b_{\mu }\text{ , \ }%
\tan \theta \equiv g^{\prime }/g  \label{gf}
\end{equation}%
($\theta $ stands for electroweak mixing angle). They acquire the masses, $%
M_{W}=gV/2$ and $M_{Z}=gV/2\cos \theta $, while an orthogonal superposition
\ of $W_{\mu }^{3}$ and$~b_{\mu }$ fields, corresponding to the
electromagnetic field%
\begin{equation}
A_{\mu }=\cos \theta ~b_{\mu }+\sin \theta \text{ }W_{\mu }^{3}~  \label{gf1}
\end{equation}%
remains massless, as usual. Then to separate the states in (\ref{ssww}) one
needs to properly shift \ the $\xi $ and $\phi $\ modes. Actually, rewriting
the mixing terms in (\ref{ssww}) in the momentum space and diagonalizing
them by the substitutions \cite{dams} \ 
\begin{equation}
\phi (k)\rightarrow \phi \left( k\right) +M_{W}\dfrac{k_{\nu }W^{\nu }\left(
k\right) }{k^{2}}\text{ , \ \ }\xi (k)\rightarrow \xi \left( k\right) -iM_{Z}%
\dfrac{k_{\nu }Z^{\nu }\left( k\right) }{k^{2}}  \label{w}
\end{equation}%
one has some transversal bilinear forms for $W$ and $Z$ bosons and the new $%
\phi (k)$ and $\xi (k)$ states%
\begin{eqnarray}
&&\left\vert -k_{\mu }\phi (k)+M_{W}\left( g_{\mu \nu }-\frac{k_{\mu }k_{\nu
}}{k^{2}}\right) W^{\nu }(k)\right\vert ^{2}+  \label{bil} \\
&&+\frac{1}{2}\left[ -ik_{\mu }\xi (k)+M_{Z}\left( g_{\mu \nu }-\frac{k_{\mu
}k_{\nu }}{k^{2}}\right) Z^{\nu }(k)\right] ^{2}  \notag
\end{eqnarray}%
to be separated. As a result, the NSM Lagrangian with the gauge fixing
conditions (\ref{ag}) included determines eventually the propagators for
massless photon and massive $W$ and $Z$ bosons in the form 
\begin{eqnarray}
D_{\mu \nu }^{(\gamma )}(k) &=&\frac{-i}{k^{2}+i\epsilon }\left( g_{\mu \nu
}-\frac{n_{\mu }k_{\nu }+k_{\mu }n_{\nu }}{(nk)}+\frac{n^{2}k_{\mu }k_{\nu }%
}{(nk)^{2}}\right) \text{ ,}  \label{prop1} \\
D_{\mu \nu }^{(W,Z)}(k) &=&\frac{-i}{k^{2}-M_{W,Z}^{2}+i\epsilon }\left(
g_{\mu \nu }-\frac{n_{\mu }k_{\nu }+k_{\mu }n_{\nu }}{(nk)}+\frac{%
n^{2}k_{\mu }k_{\nu }}{(nk)^{2}}\right) \text{ .}  \notag
\end{eqnarray}%
(where $(nk)$ stands, as usual, for a contraction $n_{\mu }k^{\mu }$).
Meanwhile, propagators for massless scalar fields $\phi $ and $\xi $ amount
to 
\begin{equation}
D^{(\phi )}(k)=\frac{i}{k^{2}}\text{ , \ \ }D^{(\xi )}(k)=\frac{i}{k^{2}}
\label{prop2}
\end{equation}%
These fields correspond to unphysical particles in a sense that they could
not appear as incoming or outgoing lines in Feynman graphs. On the other
hand, they have some virtual interactions with Higgs boson $h$, and $W$ and $%
Z$ bosons that will be taken into account when considering the corresponding
processes (see below).

Apart from the bilinear terms (\ref{ssww}), some new field bilinears appear
from \ the $n$-oriented Higgs field covariant derivative term $\left\vert
n^{\lambda }D_{\lambda }H\right\vert ^{2}$ (see below Eq. (\ref{dm})) when
the this NSM is further extended to ENSM (\ref{Ltot}). They amount to

\begin{equation}
\delta _{h}[(n_{\mu }\partial ^{\mu }h)^{2}+\left\vert n_{\mu }\partial
^{\mu }\phi \right\vert ^{2}+(n_{\mu }\partial ^{\mu }\xi )^{2}]
\end{equation}%
where $\delta _{h}=\alpha _{h}(M^{2}/M_{P}^{2})$. Inclusion of the last two
terms in the procedure of the $\phi -W$ and $\xi -Z$ separation discussed
above will change a little the form of their propagators (\ref{prop1}, \ref%
{prop2}). We do not consider this insignificant change here.

\subsection{SLIV interactions in NSM}

\subsubsection{The gauge interactions}

The Goldstone $b$-field interactions are given by the Lagrangian $\mathcal{L}%
_{NSM}$ (\ref{SM}) and particularly by its pure nonlinear part $\mathcal{L}%
_{nSM}$ (\ref{lint}) which includes in the leading order in $1/M$ the
trilinear self-interaction term of the new hypercharge vector field $b_{\mu
}=\cos \theta ~A_{\mu }-\sin \theta ~Z_{\mu }$ and, besides, the
quadrilinear couplings of this field with left-handed and right-handed
fermions, and Higgs boson. All of them have Lorentz noncovariant (preferably
oriented) form and, furthermore, they violate $CPT$ invariance as well. For
the Higgs boson part \ in $\mathcal{L}_{nSM}$ one has in the leading order
in $1/M$ using the parametrization (\ref{h})%
\begin{equation}
\mathcal{L}_{nSM}(H)=\frac{1}{2M}g^{\prime }\left( b_{\rho }\right) ^{2}%
\left[ \left( h+V\right) (n_{\mu }\partial ^{\mu })\xi -\frac{i}{2}\left[
\phi ^{\ast }(n_{\mu }\partial ^{\mu })\phi -\phi (n_{\mu }\partial ^{\mu
})\phi ^{\ast }\right] \right] \text{ .}  \label{li}
\end{equation}%
so that the quadrilinar interactions of $b_{\mu }$ field with the would-be
Goldstone bosons $\xi $ and\ $\phi (\phi ^{\ast })$ inevitably emerge. For
the properly separated $\phi -W$ and $\xi -Z$ \ states, which is reached by
the replacements (\ref{w}), there appear trilinear and quadrilinear
couplings between all particles involved in the Higgs sector (photon, $W$, $Z
$, Higgs bosons, and $\phi $ and $\xi $ fields) as directly follows from the
Lagrangian (\ref{li}) taken in the momentum space after corresponding
substitutions of (\ref{w}) and $b_{\mu }=\cos \theta ~A_{\mu }-\sin \theta
~Z_{\mu }$, respectively.

\subsubsection{Yukawa sector}

Now let us turn to the Yukava sector whose Lagrangian is

\begin{eqnarray}
\mathcal{L}_{Yuk} &=&-G\left[ \overline{L}He_{R}+\overline{e}_{R}H^{+}L%
\right] =  \label{y1} \\
&=&-\dfrac{G}{\sqrt{2}}\left[ \left( h+V\right) \overline{e}e+i\xi \overline{%
e}\gamma ^{5}e+\overline{e}_{R}\Phi ^{\ast }\nu _{l}+\overline{\nu }_{l}\Phi
e_{R}\right]   \notag
\end{eqnarray}%
Due to the $\xi $ field redifinition (\ref{w}) there appears one extra
(Yukava type) $Z$ boson coupling, which in the momentum space has the form\ 

\begin{equation}
\mathcal{L}_{Yuk}(Zee)=-\dfrac{G}{\sqrt{2}}M_{Z}\dfrac{k_{\nu }Z^{\nu
}\left( k\right) }{k^{2}}\overline{e}\gamma ^{5}e=-\dfrac{g}{2\cos \theta }%
m_{e}\dfrac{k_{\nu }Z^{\nu }\left( k\right) }{k^{2}}\overline{e}\gamma ^{5}e
\label{y2}
\end{equation}%
The similar extra coupling appears for the charged $W$ boson as well when it
is separated from the $\phi $ field due to the replacement (\ref{w}).

\subsection{Lorentz preserving SLIV\ processes}

We show now by a direct calculation of some tree level amplitudes that the
physical Lorentz invariance being intact in the massless nonlinear QED \cite%
{6, 7} is still survived in the nonlinear SM. Specifically, we will
calculate \ matrix elements of two SLIV\ processes naturally emerging in
NSM. One of them is the elastic photon-electron scattering and another is
the elastic$\ Z$\ boson scattering on an electron.

\subsubsection{Photon-electron scattering}

This process in lowest order is concerned with four diagrams one of which is
given by the direct contact photon-photon-fermion-fermion vertex generated
by the $b^{2}$-fermion-fermion coupling in (\ref{lint}), while three others
are pole diagrams where the scattered photon and fermion exchange a virtual
photon, $Z$ boson and $\xi $ field, respectively. Their vertices are given,
apart from the standard gauge boson-fermion couplings in $\mathcal{L}%
_{SM}(B_{\mu }\rightarrow b_{\mu })$ (\ref{SM}), by the SLIV $b^{3}$ and $%
b^{2}$-fermion couplings in (\ref{lint}) and by the $b^{2}$-$\xi $ coupling
in (\ref{li}), and also by Yukava couplings (\ref{y1}, \ref{y2}).

So, one has first directly from the $b^{2}$-fermion coupling the matrix
element corresponding to the contact diagram

\begin{equation}
\mathcal{M}_{c}=i\dfrac{3g}{4M}\sin \theta \cos \theta (\epsilon
_{1}\epsilon _{2})\overline{u}_{2}\gamma ^{\rho }n_{\rho }(1+\dfrac{\gamma
^{5}}{3})u_{1}  \label{c}
\end{equation}%
when expressing it through the weak isotopic constant $g$ and Weinberg angle 
$\theta $ (where $(\epsilon _{1}\epsilon _{2})$ stands for a salar product
of photon polarization vectors $\epsilon _{1\mu }$ and $\epsilon _{2\mu }$).

Using then the vertex for the ordinary SM photon-electron coupling,

\begin{equation}
-g\sin \theta \gamma ^{\mu }
\end{equation}%
together with vertex corresponding to the SLIV\ three-photon coupling,

\begin{equation}
-\ \frac{i}{M}\cos ^{3}\theta \left[ (nq)q_{\nu }g_{\lambda \rho
}+(nk_{1})k_{1\lambda }g_{\nu \rho }+(nk_{2})k_{2\rho }g_{\nu \lambda }%
\right] 
\end{equation}%
(where $k_{1,2}$ are ingoing and outgoing photon 4-momenta and $q=k_{2}-k_{1}
$, while $(nk_{1,2})$ and $(nq)$ are their contractions with the unit vector 
$n$) and photon propagator (\ref{prop1}), one comes to the matrix element
for the first pole diagram with the photon exchange

\begin{equation}
\mathcal{M}_{p1}=-i\frac{g}{M}\cos ^{3}\theta \sin \theta (\epsilon
_{1}\epsilon _{2})\overline{u}_{2}\gamma ^{\mu }n_{\mu }u_{1}  \label{p1}
\end{equation}

Analogously, combining the joint vertex for the Lorentz invariant $Z$
boson-fermion couplings which include both an ordinary SM coupling and extra
Yukava coupling (\ref{y2}) appearing due to a general parametrization (\ref%
{h}),

\begin{equation}
i\dfrac{g}{2\cos \theta }\left[ \dfrac{1}{2}\gamma ^{\mu }\left( 3\sin
^{2}\theta -\cos ^{2}\theta +\gamma ^{5}\right) -m_{e}\gamma ^{5}\dfrac{%
q_{\mu }}{q^{2}}\right]
\end{equation}%
with the vertex for the SLIV photon-photon-Z boson coupling,

\begin{equation}
i\frac{\cos ^{2}\theta \sin \theta }{M}\left[ (1-\dfrac{M_{Z}^{2}}{q^{2}}%
)(nq)q_{\nu }g_{\lambda \rho }+(nk_{1})k_{1\lambda }g_{\nu \rho
}+(nk_{2})k_{2\rho }g_{\nu \lambda }\right] \text{ },
\end{equation}%
one finds the matrix element corresponding to the second pole diagram with
the $Z$-boson exchange 
\begin{equation}
\mathcal{M}_{p2}=-i\ \frac{g}{2M}\sin \theta \cos \theta (\epsilon
_{1}\epsilon _{2})\overline{u}_{2}[\gamma ^{\mu }n_{\mu }(1-2\cos 2\theta
+\gamma ^{5})/2+\gamma ^{5}(nq)m_{e}/q^{2}]u_{1}  \label{p2}
\end{equation}%
where was also properly used Dirac equation for on-shell fermions and $Z$%
-boson propagator (\ref{prop1}).

And lastly, the third pole diagram with the $\xi $ field exchange include
two vertices, the first corresponds to Yukava $\xi ee$ coupling (\ref{y1}), 
\begin{equation}
\dfrac{g}{2\cos \theta }\dfrac{m_{e}}{M_{Z}}\gamma ^{5}
\end{equation}%
while the second to the SLIV\ $\xi $-photon-photon one (\ref{li}) 
\begin{equation}
M_{Z}\cos ^{2}\theta \sin \theta (\epsilon _{1}\epsilon _{2})(nq)
\end{equation}%
that leads, using the $\xi $ field propagator (\ref{prop2}), to the matrix
element%
\begin{equation}
\mathcal{M}_{p3}=i\dfrac{g}{2M}\dfrac{m_{e}}{q^{2}}\sin \theta \cos \theta
(\epsilon _{1}\epsilon _{2})(nq)\overline{u}_{2}\gamma ^{5}u_{1}  \label{p3}
\end{equation}%
Putting together all these contributions one can readily see that the total
SLIV\ induced matrix element for the Compton scattering taken in the lowest
order precisely vanishes,\ \ \ \ \ \ \ \ 
\begin{equation}
\mathcal{M}_{SLIV}(\gamma +e\rightarrow \gamma +e)=\mathcal{M}_{c}+\mathcal{M%
}_{p1}+\mathcal{M}_{p2}+\mathcal{M}_{p3}=0\text{ .}
\end{equation}

\subsubsection{Z boson scattering on electron}

For this process there are similar four diagrams - one is the $Z$-$Z$%
-fermion-fermion contact diagram and three others are pole diagrams where
the scattered $Z$ boson and fermion exchange a virtual photon, $Z$ boson and 
$\xi $ field, respectively. Their vertices are also given by the
corresponding couplings in the nonlinear SM Lagrangian terms (\ref{SM}, \ref%
{lint}, \ref{li}, \ref{y1}, \ref{y2}). One can readily find that the matrix
elements for the contact and pole diagrams differ from the similar diagrams
in the photon scattering case only by the Weinberg angle factor%
\begin{equation}
\mathcal{M}_{c}^{\prime }=\tan ^{2}\theta \mathcal{M}_{c}\text{ , \ \ \ }%
\mathcal{M}_{pi}^{\prime }=\tan ^{2}\theta \mathcal{M}_{pi}\text{ \ }%
(i=1,2,3)
\end{equation}%
so that we have the vanished total matrix element in this case as well%
\begin{equation}
\mathcal{M}_{SLIV}(Z+e\rightarrow Z+e)=\mathcal{M}_{c}^{\prime }+\mathcal{M}%
_{p1}^{\prime }+\mathcal{M}_{p2}^{\prime }+\mathcal{M}_{p3}^{\prime }=0\text{
.}
\end{equation}

\subsubsection{Other processes}

In the next order $O(1/M^{2})$ some new SLIV processes, such as
photon-photon, $Z$-$Z,$ photon-$Z$ boson scatterings, also appear in the
tree approximation. Their amplitudes are related, as in the above, to
photon, $Z$ boson and $\xi $ field exchange diagrams and the contact $b^{4}$
interaction diagrams following from the higher terms in $\frac{b_{\nu }^{2}}{%
M^{2}}$ in the Lagrangian (\ref{lint}). Again, all these four diagrams are
exactly cancelled giving no the physical Lorentz violating contributions.
Actually, this argumentation can be readily extended to the SLIV processes
taken in any tree-level order in $1/M$.

Most likely, a similar conclusion can be derived for SLIV loop contributions
as well. Actually, as in the massless QED case considered earlier \cite{7},
the corresponding one-loop matrix elements in NSM may either vanish by
themselves or amount to the differences between pairs of the similar
integrals whose integration variables are shifted relative to each other by
some constants (being in general arbitrary functions of external 4-momenta
of the particles involved) that in the framework of dimensional
regularization leads to their total cancellation. So, NSM not only
classically but also at quantum level appears to be physically
indistinguishable from a conventional SM. This, in turn, means that the SLIV
condition (\ref{const}) taken in the Standard Model\ is merely reduced to a
possible gauge choice for the hypercharge gauge field $B_{\mu }$, while the $%
S$-matrix remains unaltered under such a gauge convention.

\section{Extended Nonlinear Standard Model}

We now turn to the extended NSM (or ENSM) with the higher-dimensional tensor
coupling terms included, namely in some minimal form they have in equations (%
\ref{Ltot}) and (\ref{fgh}). In contrast to the PGI vector couplings which
are simply absorbed in NSM (only leading to an insignificant redifinition of
hypercharge coupling constant), these terms lead, as we show here, to the
physical SLIV with a number of specific Lorentz breaking effects appearing
through the slightly deformed dispersion relations for all SM fields
involved. Being naturally \ suppressed at low energies these effects may
become detectable in high energy physics and astrophysics. They include a
considerable change in the Greisen-Zatsepin-Kouzmin\ (GZK) cutoff for
ultra-high energy (UHE) cosmic-ray nucleons, possible stability of
high-energy pions and weak bosons and, on the contrary, instability of
photons, very significant increase of the radiative muon and kaon decays,
and some others. In this connection, the space-like Lorentz breaking
effects, due to a possible spatial anisotropy of which the current
observational limitations appear to be much weaker, may be of special
interest. Relative to the previous pure phenomenological studies \cite{2,3},
our semi-theoretical approach allows us to be more certain in predictions or
check up some ad hoc assumptions made till now.

\subsection{The basic bilinear and trilinear terms}

So, we proceed to a systematic study of the total ENSM Lagrangian (\ref{Ltot}%
). First, we express the new PGI terms in (\ref{Ltot}) through the
hypercharge NG modes $b_{\mu }$. Using again the equations (\ref{BB}) and (%
\ref{B}) in which, however, due to the high dimensionality of the tensor PGI
couplings considered, the terms of the order $O(1/M)$ are omitted, we have  
\begin{equation}
\mathcal{L}_{ENSM}=\mathcal{L}_{NSM}+\dfrac{\alpha }{M_{P}^{2}}[b_{\mu
}b_{\nu }+n^{2}(n_{\mu }b_{\nu }+n_{\nu }b_{\mu })M+n_{\mu }n_{\nu
}M^{2}]T^{\mu \nu }(f,g,h)\text{ .}  \label{len}
\end{equation}%
\ Here the total energy-momentum tensor $T^{\mu \nu }(f,g,h)$ is taken as a
sum given in (\ref{fgh}) with the corresponding ("the Lagrangian
subtracted")\ energy-momentum tensors of fermions ($T_{f}^{\mu \nu }$),
gauge fields ($T_{g}^{\mu \nu }$) and Higgs boson ($T_{h}^{\mu \nu }$),
respectively

\begin{eqnarray}
T_{f}^{\mu \nu } &=&\frac{i}{2}\left[ \overline{L}\gamma ^{\{\mu }D^{\nu
\}}L+\bar{e}_{R}\gamma ^{\{\mu }D^{\nu \}}e_{R}\right] \text{ \ },  \notag \\
T_{g}^{\mu \nu } &=&-B^{\mu \rho }B_{\rho }^{\nu }-W^{(i)\mu \rho }W_{\rho
}^{(i)\nu }\text{ ,}  \label{TT} \\
T_{h}^{\mu \nu } &=&(D^{\mu }H)^{+}D^{\nu }H+(D^{\nu }H)^{+}D^{\mu }H  \notag
\end{eqnarray}%
which all are symmetrical (in spacetime indices) and gauge invariant. One
can then use that the $W_{\mu }^{i}$ bosons ($i=1,2,3$) likewise the NG
field $b_{\mu }$ are taken in the axial gauge (\ref{ag}), due to which one
has one noticeable simplification - their preferably oriented covariant
derivatives amount to ordinary derivatives%
\begin{equation}
n_{\mu }D^{\mu }(b,W^{i})=n_{\mu }\partial ^{\mu }\text{ .}
\end{equation}%
Eventually, the total Lagrangian (\ref{Ltot}) with all leading couplings
involved comes to the sum  
\begin{equation}
\mathcal{L}_{ENSM}=\mathcal{L}_{NSM}+\mathcal{L}_{ENSM2}+\mathcal{L}_{ENSM3}
\label{Tmn}
\end{equation}%
where the nonlinear SM Lagrangian $\mathcal{L}_{NSM}$ up to the $1/M$ order
terms was discussed above (\ref{SM}, \ref{lint}), while for the new terms in
the extended Lagrangian $\mathcal{L}_{ENSM}$ we have only included the
bilinear and trilinear terms in fields involved, $\mathcal{L}_{ENSM2}$ and $%
\mathcal{L}_{ENSM3},$ respectively. Just these terms could determine the
largest deviations from a conventional SM.

Let us consider first these bilinear terms. One can readily see that they
appear from a contraction of the last term $n_{\mu }n_{\nu }M^{2}$ in the
square bracket in (\ref{len}) with energy momentum tensors $T_{f,g,h}^{\mu
\nu }$. As a result, one finally comes to the bilinear terms collected in 
\begin{eqnarray}
\mathcal{L}_{ENSM2} &=&i\delta _{f}[\overline{L}\left( \gamma ^{\mu }n_{\mu
}n_{\nu }\partial ^{\nu }\right) L+\bar{e}_{R}\left( \gamma ^{\mu }n_{\mu
}n_{\nu }\partial ^{\nu }\right) e_{R}]  \label{dm} \\
&&-\delta _{g}n_{\mu }n_{\nu }(B^{\mu \rho }B_{\rho }^{\nu }+W^{(i)\mu \rho
}W_{\rho }^{(i)\nu })+2\delta _{h}\left\vert n_{\nu }\partial ^{\nu
}H\right\vert ^{2}  \notag
\end{eqnarray}%
containing the presumably small parameters\footnote{%
Remind that the fermion parameter $\delta _{f}=\alpha _{f}M^{2}/M_{P}^{2}$
is generally different for quarks and leptons, and also depends on the
quark-lepton family considered, whereas the parameter $\delta _{g}=\alpha
_{g}M^{2}/M_{P}^{2}$ is taken to be the same for all SM gauge bosons. } $%
\delta _{f,g,h}=\alpha _{f,g,h}M^{2}/M_{P}^{2}$ since the SLIV scale $M$ is
generally proposed to be essentially lower than Planck mass $M_{P}$. These
bilinear terms modify dispersion relations for all fields involved, and
lead, in contrast to the nonlinear SM given by Lagrangian $\mathcal{L}_{NSM}$
(\ref{SM}, \ref{lint}), to the physical Lorentz violation (see below).

Let us turn now to the trilinear Lorentz breaking terms in $\mathcal{L}%
_{ENSM}$. They emerge from the contraction of the term $n^{2}(n_{\mu }b_{\nu
}+n_{\nu }b_{\mu })M$ in the square bracket in (\ref{len}) with the
energy-momentum tensors $T_{f,g,h}^{\mu \nu }$. One can see that only
contractions with derivative terms in them give the nonzero results so that
we have for the corresponding couplings for fermions%
\begin{equation}
\mathcal{L}_{ENSM3}=n^{2}\frac{\delta _{f}}{M}b_{\mu }\left[ i\overline{L}%
\left( \gamma ^{\mu }n_{\nu }\partial ^{\nu }+\gamma ^{\nu }n_{\nu }\partial
^{\mu }\right) L+i\bar{e}_{R}\left( \gamma ^{\mu }n_{\nu }\partial ^{\nu
}+\gamma ^{\nu }n_{\nu }\partial ^{\mu }\right) e_{R}\right] \text{ .}
\label{n3}
\end{equation}%
They present in fact the new type of interaction of the hypercharge
Goldstone vector field $b_{\mu }$ with the fermion matter which does not
depend on the gauge constant value $g^{\prime }$ at all. Remarkably,\ the
inclusion of other quark-lepton families into the consideration will
necessarily lead to the flavour-changing processes once the related mass
matrices of leptons and quarks are diagonalized. The point is, however, that
all these coupling in (\ref{n3}) are further suppressed by the SLIV scale $M$
and, therefore, may only become significant at superhigh energies being
comparable with this scale. In this connection, the flavour-changing
processes stemming from the less suppressed bilinear couplings (\ref{dm})
appear much more important. We will consider these processes later.

\subsection{Modified dispersion relations}

The bilinear terms collected in the Lagrangian $\mathcal{L}_{ENSM2}$ lead,
as was mentioned above, to modified dispersion relation for all fields
involved.

\subsubsection{Fermions}

Due \ to the chiral fermion content in the Standard Model we use for what
follows the chiral basis for $\gamma $ matrices 
\begin{equation}
\gamma _{\mu }=\left( 
\begin{array}{cc}
0 & \sigma ^{\mu } \\ 
\overline{\sigma }^{\mu } & 0%
\end{array}%
\right) \text{ , \ \ }\gamma _{5}=\left( 
\begin{array}{cc}
-1 & 0 \\ 
0 & 1%
\end{array}%
\right) \text{ , \ \ \ }\sigma ^{\mu }\equiv (1,\sigma ^{i})\text{ , \ \ }%
\overline{\sigma }^{\mu }\equiv (1,-\sigma ^{i})  \label{cb}
\end{equation}%
and take the conventional notations for scalar products of 4-momenta $p_{\mu
}$, unit Lorentz vector $n_{\mu }$ and four-component sigma matrices $\sigma
^{\mu }(\overline{\sigma }^{\mu })$, respectively, i.e. $p^{2}\equiv $ $%
p_{\mu }p^{\mu },$ $(np)\equiv $ $n_{\mu }p^{\mu },$ $\sigma \cdot p\equiv $ 
$\sigma _{\mu }p^{\mu }$ and $\sigma \cdot n\equiv $ $\sigma _{\mu }n^{\mu }.
$ We will discuss below Lorentz violation (in a form conditioned by the
partial gauge invariance) in the chiral basis for fermions in some detail.

\paragraph{Neutrino.}

The Lorentz noncovariant terms for neutrino and electron in $\mathcal{L}%
_{ENSM2}$ has a form\ 

\begin{equation}
i\delta _{f}\left[ \bar{\nu}(\gamma ^{\rho }n_{\rho })n^{\lambda }\partial
_{\lambda }\nu +\bar{e}_{L}(\gamma ^{\rho }n_{\rho })n^{\lambda }\partial
_{\lambda }e_{L}+\bar{e}_{R}(\gamma ^{\rho }n_{\rho })n^{\lambda }\partial
_{\lambda }e_{R}\right]   \label{2oo}
\end{equation}%
So, the modified Weyl equation for the neutrino spinor $u_{\nu }(p)$ in the
momentum space, when one assumes the standard plane-wave relation 
\begin{equation}
\nu (x)=u_{\nu }(p)\exp (-ip_{\mu }x^{\mu })\text{ \ \ \ (}p_{0}>0\text{) ,}
\label{sol1}
\end{equation}%
simply comes in the chiral basis for $\gamma $ matrices (\ref{cb}) to%
\begin{equation}
\text{\ }[(\overline{\sigma }\cdot p)+\delta _{f}(\overline{\sigma }\cdot
n)(np)]u_{\nu }(p)=0  \label{2c}
\end{equation}%
In terms of\ the new 4-momentum%
\begin{equation}
p_{\mu }^{\prime }=\ p_{\mu }+\delta _{f}(np)n_{\mu }
\end{equation}%
it acquires a conventional form

\begin{equation}
(\overline{\sigma }\cdot p^{\prime })u_{\nu }(p)=0  \label{ndr}
\end{equation}%
So, \ in terms of the "shifted" 4-momentum $p_{\mu }^{\prime }$ the neutrino
dispersion relation satisfies a standard equation $p^{\prime 2}=0$ that gives

\begin{equation}
p^{\prime 2}=p^{2}+2\delta _{f}(np)^{2}+\delta _{f}^{2}n^{2}(np)^{2}=0
\label{dr1}
\end{equation}%
while the solution for $u_{\nu }(p^{\prime }),$ as directly follows from (%
\ref{ndr}), is 
\begin{equation}
u_{\nu }(p)=\sqrt{\sigma \cdot p^{\prime }}\xi   \label{2d}
\end{equation}%
where $\xi $ is some arbitrary two-component spinor.

\paragraph{Electron.}

For electron, the picture is a little more complicated. In the same chiral
basis one has from the conventional and SLIV induced terms (\ref{2oo}) the
modified Dirac equations for the two-component left-handed and right-handed
spinors describing electron. Indeed, assuming again the standard plane-wave
relation 
\begin{equation}
e(x)=\binom{u_{L}(p)}{u_{R}(p)}\exp (-ip_{\mu }x^{\mu })\text{ , \ }p_{0}>0
\label{sol}
\end{equation}%
one comes to the equations%
\begin{eqnarray}
\text{\ }(\overline{\sigma }\cdot p^{\prime })u_{L} &=&mu_{R}\text{ \ \ }
\label{em} \\
(\sigma \cdot p^{\prime })u_{R} &=&mu_{L}\text{ \ }  \notag
\end{eqnarray}%
where we have written them in terms of 4-momenta $p^{\prime }$ 
\begin{equation}
p_{\mu }^{\prime }=p_{\mu }+\delta _{f}(np)n_{\mu }
\end{equation}%
being properly shifted in the preferred spacetime direction. Proceeding with
a standard squaring procedure one come to another pair of equations%
\begin{eqnarray}
(\sigma \cdot p^{\prime })(\overline{\sigma }\text{ }\cdot p^{\prime })u_{L}
&=&m^{2}u_{L}  \label{eqs} \\
\text{\ }(\overline{\sigma }\cdot p^{\prime })(\sigma \cdot p^{\prime
})u_{R} &=&m^{2}u_{R}  \notag
\end{eqnarray}%
being separated for left-handed and right-handed spinors. So, \ in terms of
the "shifted" 4-momentum $p_{\mu }^{\prime }$ the electron dispersion
relation satisfies a standard equation $p^{\prime 2}=m^{2}$ that gives

\begin{equation}
p^{\prime 2}=p^{2}+2\delta _{f}(np)^{2}+\delta _{f}^{2}n^{2}(np)^{2}=m^{2}
\label{f1}
\end{equation}%
while the solutions for $u_{L}(p)$ and $u_{R}(p)$ spinors in the chiral
basis taken are 
\begin{equation}
u_{L}(p)=\sqrt{\sigma \cdot p^{\prime }}\xi \text{ , \ \ }u_{R}(p^{\prime })=%
\sqrt{\overline{\sigma }\cdot p^{\prime }}\xi 
\end{equation}%
where $\xi $ is some arbitrary two-component spinor.

Further, one has to derive the orthonormalization condition for Dirac
four-spinors $u(p)=\binom{u_{L}(p)}{u_{R}(p)}$ in the presence of SLIV and
also the spin summation condition over all spin states of a physical
fermion. Let us propose first the orthonormalization condition for the
helicity eigenspinors $\xi ^{s}$ 
\begin{equation}
\xi ^{s\dagger }\xi ^{s^{\prime }}=\delta ^{ss^{\prime }}
\end{equation}%
where index $s$ stands to distinguish the "up" and "down" states. In
consequence, one has for the Hermitian conjugated and Dirac conjugated
spinors, respectively, 
\begin{equation}
u^{s\dagger }(p)u^{s^{\prime }}(p)=2[p_{0}+\delta _{f}(np)n_{0}]\delta
^{ss^{\prime }},\text{ \ }\overline{u}^{s}u^{s^{\prime }}=2m\delta
^{ss^{\prime }}  \label{ort}
\end{equation}%
Note that, whereas the former is shifted in energy $p_{0}$ for a time-like
Lorentz violation, the latter appears exactly the same as in the Lorentz
invariant theory for both the time-like and space-like SLIV.

Analogously, one has the density matrices for Dirac spinors allowing to sum
over the polarization states of a fermion. The simple calculation using the
unit "density" matrix for the generic $\xi ^{s}$ spinors%
\begin{equation}
\xi ^{s}\xi ^{s\dagger }=\left( 
\begin{array}{cc}
1 & 0 \\ 
0 & 1%
\end{array}%
\right)   \label{d}
\end{equation}%
(summation in the index $s$ is supposed) finally gives 
\begin{equation}
u^{s}(p)\text{\ }\overline{u}^{s}(p)=\left( 
\begin{array}{cc}
m & \sigma \cdot p^{\prime } \\ 
\overline{\sigma }\cdot p^{\prime } & m%
\end{array}%
\right) =\gamma ^{\mu }[p_{\mu }+\delta _{f}(np)n_{\mu }]+m  \label{dd}
\end{equation}%
when writing it in terms of the conventional Dirac $\gamma $ matrices (\ref%
{cb}).

\paragraph{Positron.}

In conclusion, consider antifermions in the SLIV extended theories. As
usual, one identifies them with the negative energy solutions. Their
equations in the momentum space appear from the plane-wave expression with
the opposite sign in the exponent%
\begin{equation}
e(x)=\binom{v_{L}(p)}{v_{R}(p)}\exp (ip_{\mu }x^{\mu })\text{ , \ }p_{0}>0
\label{soll}
\end{equation}%
and actually follow from the equations (\ref{em}), if one replaces $%
m\rightarrow -m.$ As a result, their solution have a form%
\begin{equation}
v_{L}(p)=\sqrt{\sigma \cdot p^{\prime }}\chi \text{ , \ \ }v_{R}(p)=-\sqrt{%
\overline{\sigma }\cdot p^{\prime }}\chi   \label{em1}
\end{equation}%
where $\chi $ stands for some other spinors which are related to the spinors 
$\xi ^{s}$. This relation is given, as usual, by charge conjugation $C$ 
\begin{equation}
\chi ^{s}=i\sigma _{2}(\xi ^{s})^{\ast }
\end{equation}%
where the star means the complex conjugation\footnote{%
This conforms with a general definition of the $C$ conjugation for the Dirac
spinors as an operation $u(p)^{c}=C\overline{u}(p)^{T}=i\gamma _{2}u^{\ast
}(p_{0},p_{i})$, where one identifies $u(p)^{c}=v(p),$ while $C$ matrix is
chosen as $i\gamma _{0}\gamma _{2}$.}. This form of $\chi ^{s}$ says that
this operation actually interchanges the "up" and "down" spin states given
by $\xi ^{s}.$ All other equations for positron states described by the
corresponding four-spinors $v^{s}(p)$, namely those for the normalization 
\begin{equation}
v^{s\dagger }(p)v^{s^{\prime }}(p)=2[p_{0}+\delta _{f}(np)n_{0}]\delta
^{ss^{\prime }},\text{ \ }\overline{v}^{s}v^{s^{\prime }}=-2m\delta
^{ss^{\prime }}
\end{equation}%
and density matrices%
\begin{equation}
u^{s}(p)\text{\ }\overline{u}^{s}(p)=\left( 
\begin{array}{cc}
-m & \sigma \cdot p^{\prime } \\ 
\overline{\sigma }\cdot p^{\prime } & -m%
\end{array}%
\right) =\gamma ^{\mu }[p_{\mu }+\delta _{f}(np)n_{\mu }]-m\text{ ,}
\end{equation}%
also straightforwardly emerge. One can notice, that, apart from the standard
sign changing before the mass term all these formulas are quite similar to
the corresponding expressions in the positive energy solution case.

\subsubsection{Gauge bosons}

To establish the form of modified dispersion relations for gauge fields one
should take into account, apart from their standard kinetic terms in the
NSM\ Lagrangian (\ref{SM}), the\ quadratic terms appearing from SLIV (\ref%
{dm}). \ 

\paragraph{Photon.}

\ \ Let us consider first the photon case. The modification of photon
kinetic term appears from the modifications of kinetic terms for $B$ and $%
W^{3}$ gauge fields both taken in the axial gauge. These terms, due to the
invariant quadratic form of the "bilinear" SLIV Lagrangian $\mathcal{L}%
_{ENSM2}$ (\ref{dm}), lead to similar modifications for photon (\ref{gf1})
and $Z$ boson (\ref{gf}). So, constructing kinetic terms for the photon in
the momentum space one readily finds its modified dispersion relation 
\begin{equation}
k^{2}+2\delta _{g}(nk)^{2}=0\text{ , \ \ }\delta _{g}=\alpha
_{g}(M^{2}/M_{P}^{2})  \label{ds}
\end{equation}%
while its SLIV modified propagator has a form%
\begin{equation}
D^{\mu \nu }=\frac{-i}{k^{2}+2\delta _{g}(nk)^{2}+i\epsilon }\left[ g^{\mu
\nu }-\frac{1}{1+2\delta _{g}n^{2}}\left( \frac{n_{\mu }k_{\nu }+k_{\mu
}n_{\nu }}{(nk)}-n^{2}\frac{k_{\mu }k_{\nu }}{(nk)^{2}}+2\delta _{g}n_{\mu
}n_{\nu }\right) \right]   \label{pro.1}
\end{equation}%
This satisfies the conditions 
\begin{equation}
n^{\mu }D^{\mu \nu }=0\ \ ,\ \ \ \ \ \ \ k^{\mu }D^{\mu \nu }=0\ \ \ \ \ \ \
\   \label{transv.}
\end{equation}%
where the transversality condition in (\ref{transv.}) \ is imposed on the
photon "mass shell" which is now determined by the modified dispersion
relation (\ref{ds}). Clearly, \ in the Lorentz invariance limit ($\delta
_{g}\rightarrow 0$) \ the propagator (\ref{pro.1}) goes into\ the standard
propagator taken in an axial gauge (\ref{prop1}).

\paragraph{W and Z bosons.}

Analogously, constructing the kinetic operators for the massive vector
bosons one has the following modified dispersion relations for them 
\begin{equation}
k^{2}+2\delta _{g}(nk)^{2}=M_{Z,W}^{2}
\end{equation}%
To make the simultaneous modification of their propagators, one also should
take into account the terms emerged from the Higgs sector. These terms
appear when, through the proper diagonalization, the Higgs bilinears
decouple from those of the massive $W$ and $Z$ bosons. Due to their
excessive length we do not present their modified propagators here.

\subsubsection{Higgs boson}

For Higgs boson (with 4-momentum $k_{\mu }$ and mass $\mu _{h}$), we have
from the properly modified Klein-Gordon equation, \ appearing from its basic
Lagrangian (\ref{ssww}) taken together with the last term in $\mathcal{L}%
_{ENSM2}$ (\ref{dm}), the dispersion relation 
\begin{equation}
k^{2}+2\delta _{h}(nk)^{2}=\mu _{h}^{2}\text{ , \ }\delta _{h}=\alpha
_{h}(M^{2}/M_{P}^{2})\text{ .}  \label{h11}
\end{equation}

\subsection{Lorentz breaking SLIV processes}

We are ready now to consider the SLIV contributions into some physical
processes. They include as ordinary processes where the Lorentz violation
gives only some corrections, being quite small at low energies but
considerably\ increasing with energy, so the new processes being entirely
determined by SLIV in itself. Note that the most of these processes were
considered hitherto \cite{2, 3} largely on the pure phenomenological ground.
We discuss them here in the ENSM\ framework which contains only four
effective SLIV parameters$^{10}$ $\delta _{f}$ ($f=q,l$), $\delta _{g}$ and $%
\delta _{h}$ rather than a variety of phenomenological parameters introduced
for each particular process individually \cite{3}. Indeed, one (or, at most,
two) more fermion parameters should be added in our case too when different
quark-lepton families and related flavor-changing processes are also
considered.

Another important side of our consideration is that for every physical
process we take into account, together with the direct contributions of the
SLIV couplings in \ the Lagrangian, the Lorentz violating contributions
appearing during the integration over the phase space. The latter for the
most considered processes is still actually absent in the literature.
Specifically, for decay processes, we show that when there are identical
particles (or particles belonging to the \ same quark-lepton family) in
final states one can directly work with their SLIV shifted 4-momenta (see,
for example (\ref{f1})) for which the standard dispersion relations hold
and, therefore, the standard integration over the phase space can be carried
out. At the same time, for the decaying particles by themselves the special
SLIV influenced quantity called the "effective mass" may be introduced.
Remarkably, all such decay rates in the leading order in the SLIV\ $\delta $%
-parameters are then turned out to be readily expressed in terms of the
standard decay rates, apart from that the masses of decaying particles are
now replaced by their "effective masses".

Our calculations confirm that there are lots of the potentially sensitive
tests of the Lorentz invariance, especially at superhigh energies $%
E>10^{18}eV$ \ that is an active research area for the current cosmic-ray
experiments \cite{Oge}. Some of them will be considered in detail below.

\subsubsection{Higgs boson decay into fermions}

We start by calculating the Higgs boson decay rate into an electron-positron
pair. The vertex for such process is given by the Yukawa coupling

\begin{equation}
\frac{G}{\sqrt{2}}h\overline{e}e
\end{equation}%
with the coupling constant $G$. Properly squaring the corresponding matrix
element with the electron and positron solutions given above (\ref{sol}, \ref%
{soll}) one has

\begin{eqnarray}
\left\vert \mathcal{M}_{he\overline{e}}\right\vert ^{2} &=&\frac{G^{2}}{2}%
\left( Tr[(p_{\mu }^{\prime }\gamma ^{\mu })(q_{\nu }^{\prime }\gamma ^{\nu
})]-4m^{2}\right)   \label{mat} \\
&=&2G^{2}(p_{\mu }^{\prime }q^{\prime \mu }-m^{2})  \notag
\end{eqnarray}%
where $\ p_{\mu }^{\prime }$ $\ $and$\ \ q_{\mu }^{\prime }$ \ are the SLIV\
shifted 4-momenta of electron and positron, respectively, defined as 
\begin{eqnarray}
p_{\mu }^{\prime } &=&\ p_{\mu }+\delta _{f}(np)n_{\mu }\text{ , \ \ \ \ }%
p_{\mu }^{\prime 2}=m^{2}  \label{p11} \\
q_{\mu }^{\prime } &=&\ q_{\mu }+\delta _{f}(nq)n_{\mu }\text{ , \ \ \ \ }%
q_{\mu }^{\prime 2}=m^{2}\text{ .}  \notag
\end{eqnarray}%
We use then the conservation law for the original 4-momenta of Higgs boson
and fermions 
\begin{equation}
k_{\mu }=p_{\mu }+q_{\mu }
\end{equation}%
since just these 4-momenta rather than their SLIV\ shifted 4-momenta (for
which the above conservation law only approximately works) still determine
the spacetime evolution of all freely propagating particles involved .
Rewriting this relation as 
\begin{equation}
k_{\mu }+\delta _{f}(nk)n_{\mu }=p_{\mu }^{\prime }+q_{\mu }^{\prime }
\end{equation}%
and squaring it one has, using the relations (\ref{h11}) and (\ref{p11}), 
\begin{equation*}
p_{\mu }^{\prime }q^{\prime \mu }=\mu _{h}^{2}/2-(\delta _{h}-\delta
_{f})(nk)-m^{2}
\end{equation*}%
that finally gives for the matrix element (\ref{mat}) 
\begin{equation}
\left\vert \mathcal{M}_{he\overline{e}}\right\vert ^{2}=G^{2}(\mathbf{\mu }%
_{h}^{2}-4m^{2})
\end{equation}%
where we have denoted by $\mathbf{\mu }_{h}^{2}$ the combination 
\begin{equation}
\mathbf{\mu }_{h}^{2}=\mu _{h}^{2}-2(\delta _{h}-\delta _{f})(nk)^{2}\text{ .%
}  \label{ef}
\end{equation}%
This can be considered as an "effective" mass square of Higgs boson which
goes to the standard value $\mu _{h}^{2}$ in the Lorentz invariance limit.
One can also introduce the corresponding 4-momentum 
\begin{equation}
k_{\mu }^{\prime }=k_{\mu }+\delta _{f}(nk)n_{\mu }\text{ , \ }k_{\mu
}^{\prime 2}=\mathbf{\mu }_{h}^{2}  \label{h22}
\end{equation}%
which differs from the 4-momentum determined due the Higgs boson dispersion
relation (\ref{h11}).

So, Lorentz violation due to the matrix element is essentially presented in
the "effective" mass of the decaying Higgs particle. Let us turn now to the
SLIV part stemming from an integration over the phase space of the fermions
produced. It is convenient to come from the deformed original 4-momenta ($%
k_{\mu },$ $p_{\mu },$ $q_{\mu }$) to the shifted ones ($k_{\mu }^{\prime },$
$p_{\mu }^{\prime },$ $q_{\mu }^{\prime }$) for which fermions have normal
dispersion relations given in (\ref{p11}). Actually, possible corrections to
such momentum replacement are quite negligible\footnote{%
Actually, there is the following correspondence between shifted and original
momenta when integrating over the phase space: for the delta functions this
is \ $\delta ^{4}(k^{\prime }-p^{\prime }-q^{\prime })=(1+\delta
_{f})^{-1}\delta ^{4}(k-p-q)$ (for both time-like and space-like SLIV),
while for the momentum differentials there are $\frac{d^{3}p^{\prime
}d^{3}q^{\prime }}{k_{0}^{\prime }p_{0}^{\prime }q_{0}^{\prime }}=(1+\delta
_{f})^{-3}\frac{d^{3}pd^{3}q}{k_{0}p_{0}q_{0}}$ (time-like SLIV) and $\frac{%
d^{3}p^{\prime }d^{3}q^{\prime }}{k_{0}^{\prime }p_{0}^{\prime
}q_{0}^{\prime }}=(1+\delta _{f})^{2}\frac{d^{3}pd^{3}q}{k_{0}p_{0}q_{0}}$
(space-like SLIV). So, one can use in a good approximation the shifted
momentum variables instead of the original ones.} as compared to the Lorentz
violations stemming from the "effective" mass (\ref{ef}) where they are
essentially enhanced by the factor\ $(nk)^{2}$. Actually, writing the Higgs
boson decay rate in the shifted 4-momenta we really come to a standard case,
apart from that the Higgs boson mass is now replaced by its "effective" mass
(see below). So, for this rate we still have 
\begin{equation}
\Gamma _{he\overline{e}}=\frac{G^{2}(\mathbf{\mu }_{h}^{2}-4m^{2})}{32\pi
^{2}k_{0}^{\prime }}\int \frac{d^{3}p^{\prime }d^{3}q^{\prime }}{%
p_{0}^{\prime }q_{0}^{\prime }}\delta ^{4}(k^{\prime }-p^{\prime }-q^{\prime
})\text{ .}  \label{ww}
\end{equation}%
Normally, in a standard Lorentz-invariant case this phase space integral
comes to $2\pi $. Now, for the negligible fermion (electron) mass, $\mathbf{%
\mu }_{h}^{2}>>m^{2}$ (or more exactly $\delta _{f}k_{0}^{2}>>m^{2}$), one
has using the corresponding energy-momentum relations of particles involved, 
\begin{equation}
\int \frac{d^{3}p^{\prime }d^{3}q^{\prime }}{p_{0}^{\prime }q_{0}^{\prime }}%
\delta ^{4}(k^{\prime }-p^{\prime }-q^{\prime })\simeq 2\pi \frac{%
k_{0}^{\prime }}{\sqrt{\mathbf{\mu }_{h}^{2}}}
\end{equation}%
that for Higgs boson rate eventually gives

\begin{equation}
\Gamma _{he\overline{e}}\simeq \frac{G^{2}}{16\pi }\sqrt{\mathbf{\mu }%
_{h}^{2}}\simeq \Gamma _{he\overline{e}}^{0}\left[ 1-(\delta _{h}-\delta
_{f})\frac{(nk)^{2}}{\mu _{h}^{2}}\right]   \label{h.d.r}
\end{equation}%
The superscript "$0$" in the decay rate $\Gamma $ here and below belong to
its value in the Lorentz invariance limit. Obviously, the SLIV deviation
from this value at high energies depends on a difference of delta
parameters. In the time-like SLIV case for energies $k_{0}>\mu
_{h}/\left\vert \delta _{h}-\delta _{f}\right\vert ^{1/2}$ this decay
channel breaks down, though other channels like as $h\rightarrow 2\gamma $
(or $h\rightarrow 2$ gluons) may still work if the corresponding kinematical
bound $\mu _{h}/\left\vert \delta _{h}-\delta _{g}\right\vert ^{1/2}$ for
them is higher. For the space-like SLIV the effective delta parameter
becomes dependent on the orientation of momentum of initial particle as
well, and if, for example, $\vartheta $ is the angle between \ $%
\overrightarrow{k}$ \ and \ $\overrightarrow{n}$ , \ the threshold energy is
given by \ $k_{0}>\mu _{h}/\left\vert (\delta _{h}-\delta _{f})\cos
^{2}\vartheta \right\vert ^{1/2}$. \ So, the decay rate may acquire a strong
spatial anisotropy at ultra-high energies corresponding to standard
short-lived Higgs bosons in some directions and, at the same time, to
unusually long-lived bosons in other ones.

\subsubsection{Weak boson decays}

Analogously, one can readily write the $Z$ and $W$ boson decay rates into
fermions replacing in standard formulas the $Z$ and $W$ boson masses by
their "effective masses" which similar to (\ref{ef}) are given by 
\begin{equation}
\ \mathbf{M}_{Z,W}^{2}\simeq M_{Z,W}^{2}-2(\delta _{g}-\delta _{f})(nk)^{2}
\end{equation}%
Therefore, for the $Z$ boson decay into a neutrino-antineutrino pair one has
again the factorized expression in terms of the Lorentz invariant and SLIV
contributions%
\begin{equation}
\Gamma _{Z\nu \overline{\nu }}\simeq \dfrac{g^{2}}{96\pi \cos ^{2}\theta }%
\sqrt{\mathbf{M}_{Z}^{2}}\simeq \Gamma _{Z\nu \overline{\nu }}^{0}\text{ }%
\left[ 1-(\delta _{g}-\delta _{f})\frac{(nk)^{2}}{M_{Z}^{2}}\right] 
\end{equation}%
For the $Z$ decay into massive fermions (with masses $m<<\sqrt{\mathbf{M}%
_{Z}^{2}}$) one has a standard expression though with the "effective" $Z$
boson mass square $\mathbf{M}_{Z}^{2}$ inside rather than an ordinary mass
square $M_{Z}^{2}$

\begin{equation}
\Gamma _{Zee}=\dfrac{g^{2}(1+r)}{96\pi \cos ^{2}\theta }\sqrt{\mathbf{M}%
_{Z}^{2}}=\Gamma _{Zee}^{0}\left[ 1-(\delta _{g}-\delta _{f})\frac{(nk)^{2}}{%
M_{Z}^{2}}\right] 
\end{equation}%
where, for certainty, we have focused on the decay into an electron-positron
pair and introduced, as usual, the electroweak mixing angle factor with $%
r=-4\sin ^{2}\theta \cos 2\theta $. One can see that in the leading order in 
$\delta $-parameters\ the relation between the total decay rates $\Gamma _{Ze%
\overline{e}}$ and $\Gamma _{Z\nu \overline{\nu }}$ remains the same as in
the Lorentz invariant case.

As to the conventional $W$ boson decay into an electron-neutrino pair, one
can write in a similar way%
\begin{equation*}
\Gamma _{We\nu }\simeq \dfrac{g^{2}}{48\pi }\sqrt{\mathbf{M}_{W}^{2}}\simeq
\Gamma _{We\nu _{e}}^{0}\left[ 1-(\delta _{g}-\delta _{f})\frac{(nk)^{2}}{%
M_{W}^{2}}\right] 
\end{equation*}

So, again as was in the Higgs boson case, the $Z$ and $W$ bosons at energies 
$k_{0}>M_{Z,W}/\sqrt{\delta _{g}-\delta _{f}}$ tend to be stable for the
time-like SLIV or decay anisotropically for the space-like {}one.

\subsubsection{Photon decay into electron-positron pair}

Whereas the above mentioned decays could contain some relatively small SLIV
corrections, the possible photon decay, which we now turn to, is entirely
determined by the Lorentz violation. Indeed, while physical photon remain
massless, its "effective" mass, caused by SLIV, may appear well above of the
double electron mass that kinematically allows this process to go.

The basic electromagnetic vertex for fermions in SM is given, as usual

\begin{equation}
-(ie)\text{ }\overline{e}\epsilon _{\mu }\gamma ^{\mu }e
\end{equation}%
where we denoted electric charge by the same letter $e$ as the electron
field variable $e(x)$ and introduced the photon polarization vector $%
\epsilon _{\mu }(s)$. The fermion dispersion relations in terms of the SLIV\
shifted 4-momenta and the photon "effective" mass have the form (similar to
those in the above cases) 
\begin{equation}
\ p_{\mu }^{\prime 2}=q_{\mu }^{\prime 2}=m^{2}\text{ , \ \ \ \ \ \ }\ 
\mathbf{M}_{\gamma }^{2}\simeq 2\left( \delta _{f}-\delta _{g}\right) \left(
n^{\mu }k_{\mu }\right) ^{2}\equiv \ k_{\mu }^{\prime 2}
\end{equation}%
Consequently, for the square of the matrix element one has

\begin{equation}
\left\vert \mathcal{M}_{\gamma e\overline{e}}\right\vert ^{2}=4e^{2}\left[
2(p^{\prime }\epsilon )(q^{\prime }\epsilon )-\epsilon _{\mu
}^{2}(m^{2}+(p^{\prime }q^{\prime }))\right] 
\end{equation}%
Due to the energy-momentum conservation allowing to replace

\begin{equation}
p_{\mu }^{\prime }q_{\nu }^{\prime }\rightarrow \frac{1}{12}\left( (\mathbf{M%
}_{\gamma }^{2}-4m^{2})g_{\mu \nu }+2(\mathbf{M}_{\gamma }^{2}+2m^{2})\frac{%
k_{\mu }^{\prime }k_{\nu }^{\prime }}{\mathbf{M}_{\gamma }^{2}}\right) 
\end{equation}%
\ \ and the summation over the photon polarization{\LARGE \ }states which
according to the modified photon propagator (\ref{pro.1}) has the form

\begin{equation}
\epsilon _{\mu }(s)\epsilon _{\nu }(s)=-g^{\mu \nu }+\frac{1}{1+2\delta
_{g}n^{2}}\left( \frac{n_{\mu }k_{\nu }+k_{\mu }n_{\nu }}{(nk)}-n^{2}\frac{%
k_{\mu }k_{\nu }}{(nk)^{2}}+2\delta _{g}n_{\mu }n_{\nu }\right) 
\end{equation}%
one finally comes to the properly averaged square of the matrix element

\begin{equation}
\left\vert \overline{\mathcal{M}}_{\gamma e\overline{e}}\right\vert ^{2}=%
\frac{4e^{2}}{3}(\mathbf{M}_{\gamma }^{2}+2m^{2})
\end{equation}%
Therefore, for a calculation of the photon decay rate there is only left an
integration over phase space

\begin{equation}
\Gamma _{\gamma e\overline{e}}=\frac{e^{2}}{24\pi ^{2}k_{0}^{\prime }}(%
\mathbf{M}_{\gamma }^{2}+2m^{2})\int \frac{d^{3}p^{\prime }d^{3}q^{\prime }}{%
p_{0}^{\prime }q_{0}^{\prime }}\delta ^{4}(k^{\prime }-p^{\prime }-q^{\prime
})
\end{equation}%
that in a complete analogy with the above Higgs boson decay case (\ref{ww})
leads in the limit $\mathbf{M}_{\gamma }^{2}>>m^{2}$ to the simple answer 
\begin{equation}
\Gamma _{\gamma e\overline{e}}\simeq \frac{e^{2}}{12\pi }\sqrt{\mathbf{M}%
_{\gamma }^{2}}\simeq \frac{e^{2}}{12\pi }\sqrt{2\left\vert \delta
\right\vert }k_{0}
\end{equation}%
where \ $\delta =\delta _{f}-\delta _{g}$ \ for the time-like violation and
\ \ $\delta =(\delta _{f}-\delta _{g})\cos ^{2}\vartheta $ \ for the
space-like one with an angle $\vartheta $ between the preferred SLIV
direction and the starting photon 3-momentum. Note that, though, as was
indicated in \cite{3}, the detection of the primary cosmic-ray photons with
energies up to $20$ $TeV$ sets the stringent limit on the Lorentz violation,
this limit belongs in fact to the time-like SLIV case giving $\left\vert
\delta _{f}-\delta _{g}\right\vert <10^{-15}$ rather than to the space-like
one which in some directions may appear much more significant.

\subsubsection{Radiative muon decay}

In contrast, the muon decay process $\mu \rightarrow e+\gamma $, though
being kinematically allowed, is strictly forbidden in the ordinary SM and is
left rather small even under some of its known extensions. However, the
Lorentz violating interactions in our model may lead to the significant
flavor-changing processes both in lepton and quark sector. Particularly,
they may raise the radiative muon decay rate up to its experimental upper
limit $\Gamma _{\mu e\gamma }<10^{-11}\Gamma _{\mu e\nu \overline{\nu }}$.
The point is that the "effective" mass eigenstates of high-energy fermions
do not in general coincide with their ordinary mass eigenstates. So, if we
admit that, while inside of the each family all fermions are proposed to
have equal \ SLIV $\delta $-parameters, the different families could have in
general the different ones, say, $\delta _{e},\delta _{\mu }$ and $\delta
_{\tau }$ for the first, second and third family, respectively. As a result,
diagonalization of the fermion mass matrices will then cause small
non-diagonalities in the energy-dependent part of the fermion bilinears
presented in the $\mathcal{L}_{ENSM2}$ (\ref{dm}), even if initially they
are taken diagonal.

Let us consider, as some illustration, the electron-muon system ignoring for
the moment possible mixings of electrons and muons with tau leptons. To this
end, the Lagrangian $\mathcal{L}_{ENSM2}$ is supposed to be extended so as
to include the muon bilinears as well. Obviously, the leading diagrams
contributing into the $\mu \rightarrow e+\gamma $ are in fact two simple
tree diagrams where muon emits first photon and then goes to electron due to
the "Cabibbo rotated" bilinear couplings (\ref{dm}) or, on the contrary,
muon goes first to electron and then emits photon. Let us ignore this time
the pure kinematical part of the SLIV\ contribution following from the
deformed dispersion relations of all particles involved thus keeping in mind
only its "Cabibbo rotated" part in the properly extended bilinear couplings (%
\ref{dm}). In this approximation the radiative muon decay rate is given by 
\begin{equation}
\Gamma _{\mu e\gamma }=\frac{e^{2}}{32\pi }\frac{(pn)^{3}}{m_{\mu }^{2}}%
(\delta _{\mu }-\delta _{e})^{2}\sin ^{2}2\varphi   \label{mu}
\end{equation}%
where $p$ is the muon 4-momentum and $\varphi $ is the corresponding mixing
angle of electron and muon. Taking for their starting mass matrix $m_{ab}$
the Hermitian matrix with a typical $m_{11}=0$ texture form \cite{fr}%
\begin{equation}
m_{ab}=\left( 
\begin{array}{cc}
0 & b \\ 
b & c%
\end{array}%
\right) \text{ ,}
\end{equation}%
one has%
\begin{equation*}
\sin ^{2}2\varphi =4\frac{m_{e}}{m_{\mu }}\text{ .}
\end{equation*}%
So, though the decay rate (\ref{mu}) is in fact negligibly small when muon
is at rest, this rate increases with the cube of the muon energy and becomes
the dominant decay mode at sufficiently high energies. If we admit that
there are still detected the UHE primary cosmic ray muons possessing
energies around $10^{19}eV$ \cite{Oge} the following upper limit for the
SLIV parameters stems 
\begin{equation}
\left\vert \delta _{\mu }-\delta _{e}\right\vert <10^{-24}
\end{equation}%
provided that the branching ratio $\Gamma _{\mu e\gamma }/\Gamma _{\mu e\nu 
\overline{\nu }}$ at these energies is\ taken to be of the order one or so.
This suggests, as one can see, a rather sensitive way of observation of a
possible Lorentz violation through the search for a lifetime anomaly of
muons at ultra-high energies.

\subsubsection{The GZK cutoff revised}

One of the most interesting examples where a departure from Lorentz
invariance can essentially affect a physical process is the transition $%
p+\gamma \rightarrow \Delta $ which underlies the Greisen-Zatsepin-Kouzmin
cutoff for UHE cosmic rays \cite{gzk}. According to this idea primary
high-energy nucleons ($p$ ) should suffer an inelastic impact with cosmic
background photons ($\gamma $) due to the resonant formation of \ the first
pion-nucleon resonance $\Delta (1232)$, so that nucleons with energies above 
$\sim 5\cdot 10^{19}eV$ could not reach us from further away than $\sim 50$ $%
Mpc$. During the last decade there were some serious indications \cite{GZK}
that the primary cosmic-ray spectrum extends well beyond the GZK cutoff,
though presently the situation is somewhat unclear due to a certain
criticism of these results and new data that recently appeared \cite{Oge}.
However, no matter how things will develop, we could say that according to
the modified dispersion relations of all particles involved the GZK cutoff
will necessarily be changed at superhigh energies, if Lorentz violation
occurs.

Actually, one may expect that the modified dispersion relations for quarks
will, in turn, change dispersion relations for composite hadrons (protons,
neutrons, pions, $\Delta $ resonances etc.) depending on a particular
low-energy QCD dynamics appearing in each of these states. In general, one
could accept that their dispersion relations have the same form (\ref{f1})
as they have for elementary fermions, only their SLIV$\ \delta $ parameters
values may differ. So, for the proton and $\Delta $ there appear equations,

\begin{equation}
P_{p,\Delta }^{2}=m_{p,\Delta }^{2}-2\delta _{p,\Delta }\left( nP_{p,\Delta
}\right) ^{2}=\mathbf{m}_{p,\Delta }^{2}
\end{equation}%
respectively, which determine their deformed dispersion relations and
corresponding "effective" masses (where $P_{\mu }=(E,P_{i})$ stands for the
associated 4-momenta). To proceed, one must replace the fermion masses in a
conventional proton threshold energy for the above mentioned process%
\begin{equation}
E_{p}\geq \frac{m_{\Delta }^{2}-m_{p}^{2}}{4\omega }  \label{pr}
\end{equation}%
by their "effective" masses $\mathbf{m}_{p,\Delta }^{2}$. The target photon
energies $\omega $ in (\ref{pr}) are vanishingly small ($\omega \sim
10^{-4}eV$) and, therefore, its SLIV induced "effective mass" can be ignored
that gives an approximate equality of the fermion energies, $E_{\Delta
}=E_{p}+\omega \cong E_{p}$. As a result, the modified threshold energy for
the UHE proton scattering on the background photon via the intermediate $%
\Delta $ particle production is happened to be

\begin{equation}
E_{p}\geq \frac{m_{\Delta }^{2}-m_{p}^{2}}{2\omega +\sqrt{4\omega
^{2}+2(\delta _{\Delta }-\delta _{p})(m_{\Delta }^{2}-m_{p}^{2})}}
\label{ee}
\end{equation}

Obviously,\ if there is time-like Lorentz violation and, besides, $\delta
_{p}-\delta _{\Delta }>2\omega ^{2}/(m_{\Delta }^{2}-m_{p}^{2})$ \ this
process, as follows from (\ref{ee}), becomes kinematically forbidden at all
energies. For other values of $\delta $ parameters one could significantly
relax the GZK cutoff. The more interesting picture seems to appear for the
space-like SLIV with \ \ $\delta _{p}-\delta _{\Delta }>2\omega ^{2}/\left(
m_{\Delta }^{2}-m_{p}^{2}\right) \cos ^{2}\vartheta $ \ , where $\vartheta $
is the angle between the initial proton 3-momentum and preferred SLIV\
direction fixed by the unit vector $\overrightarrow{n}$. Actually, one could
generally observe different cutoffs for different directions, or not to have
them at all for some other directions thus \ permitting the UHE cosmic-ray
nucleons to travel over cosmological distances.

\subsubsection{Other hadron processes}

Some other hadron processes, like as the pion or nucleon decays, studied
phenomenologically earlier \cite{3} are also interesting to be reconsidered
in our semi-theoretical framework. Departures from Lorentz invariance can
significantly modify the rates of allowed hadron processes, such as $\pi
\rightarrow 
\mu
+\nu $ and $\pi \rightarrow 2\gamma ,$ at supehigh energies. In our model
these rates can be readily written replacing the mass of the decaying pion
by its "effective masses" being determined independently for each of these
cases. So, one has them again in the above mentioned factorized forms (in
the leading order in $\delta $ parameters) as%
\begin{eqnarray}
\Gamma _{\pi 
\mu
\nu } &\simeq &\Gamma _{\pi 
\mu
\nu }^{0}\left[ 1-(\delta _{\pi }-\delta _{f})\frac{(nk)^{2}}{m_{\pi }^{2}}%
\right]  \\
\Gamma _{\pi \gamma \gamma } &\simeq &\Gamma _{\pi \gamma \gamma }^{0}\left[
1-3(\delta _{\pi }-\delta _{g})\frac{(nk)^{2}}{m_{\pi }^{2}}\right]   \notag
\end{eqnarray}%
where we have used that their standard decay rates are proportional to the
first and third power of the pion mass, respectively. Therefore, the charged
pions at energies $k_{0}>m_{\pi }/\sqrt{\delta _{\pi }-\delta _{f}}$ and
neutral pions at energies $k_{0}>m_{\pi }/\sqrt{3(\delta _{\pi }-\delta _{g})%
}$ may become stable for the time-like SLIV or decay anisotropically for the
space-like {}one. As was indicated in \cite{3}, even for extremely small $%
\delta $-parameters of the order $10^{-24}\div 10^{-22}$ this phenomenon
could appear for the presently studied UHE primary cosmic ray pions
possessing energies around $10^{19}eV$ and higher.

As in the lepton sector, there also could be the SLIV induced non-diagonal
transitions in the quark sector leading to the flavor-changing processes for
hadrons. The SLIV induced radiative quark decay $s\rightarrow d+\gamma $ is
of special interest. This could make the radiative hadron decays $%
K\rightarrow \pi +\gamma $ and $\Sigma (\Lambda )\rightarrow N+\gamma $ to
become dominant at ultra-high energies just as it is for the radiative muon
decay mentioned above. Again, an absence of kaons and hyperons at these
energies or a marked decrease of their lifetime could point to the fact that
Lorentz invariance is essentially violated. 

\section{Conclusion}

We found it conceivable that an exact gauge invariance may disable some
generic features of the Standard Model which could otherwise manifest
themselves at high energies. In this connection, we have proposed the
partial gauge invariance (or PGI) in SM (\ref{ntt}) and found an appropriate
minimal form for PGI (\ref{Ltot}). This form depends on the way SLIV is
realized in SM and a special role which may play the basic Noether currents,
namely, the total hypercharge current and the total energy-momentum tensor
in the partially gauge invariant SM. These currents  and nonlinear Lorentz
realization taken together are precisely the ingredients which appeared
essential for our consideration. Just they provide the minimal PGI principle
to be reasonably well defined at least at the classical level, as was argued
in section 2.

In regard to the theory obtained, we showed first that in the simplest
nonlinear SM extension (NSM) with the vector field "length-fixing"
constraint $B_{\mu }^{2}=n^{2}M^{2}$\ the spontaneous Lorentz violation
actually holds (as it normally takes place for internal symmetries in any
nonlinear $\sigma $-model type theory) due to which the hypercharge gauge
field is converted into a vector Goldstone boson which having been then
mixed with a neutral $W^{3}$ boson of $SU(2)$ leads, as usual, to the
massless photon and massive $Z$ boson. However, in sharp contrast to an
internal symmetry case, all observational SLIV effects in NSM are turned out
to be exactly cancelled due to some remnant gauge invariance that is still
left in the theory\footnote{%
Remarkably, a similar nonlinear $\sigma $-model type modification of
conventional Yang-Mills theories and gravity with the "length-fixing"
constraint put on gauge fields appears again insufficient to lead to an
actual physical Lorentz violation \cite{jej1}.}. The point is that the SLIV
pattern according to which just the vector field (rather than some scalar
field derivative \cite{ark} or vector field stress-tensor \cite{ur})
develops the vacuum expectation value, taken as $B_{\mu }(x)=b_{\mu
}(x)+n_{\mu }M,$ may be treated in itself as a pure gauge transformation
with gauge function linear in coordinates, $\omega (x)=$ $(n_{\mu }x^{\mu })M
$. In this sense, the starting gauge invariance in SM, even being partially
broken by the nonlinear field constraint, \ leads to the conversion of SLIV
into gauge degrees of freedom of the massless NG boson $b_{\mu }(x)$. This
is what one could refer to as the generic non-observability of SLIV in a
conventional SM. Furthermore, as was shown some time ago \cite{cfn}, gauge
theories, both Abelian and non-Abelian, can be obtained by themselves from
the requirement of the physical non-observability of SLIV, caused by the
Goldstonic nature of vector fields, rather than from the standard gauge
principle.

However, a clear signal of the physical Lorentz violation inevitably appears
when one goes beyond NSM to include as well the\ higher dimensional tensor
couplings in (\ref{ntt}) that leads to the extended nonlinear SM (ENSM).
These couplings according to the minimal PGI are proposed to be determined
solely by the total energy-momentum tensor of all SM fields involved. So,
the lowest order ENSM which conforms with the chiral nature of SM and all
accompanying global and discrete symmetries, is turned out to include the
dimension-6 couplings\footnote{%
Note that in the pure QED with vectorlike (rather than chiral) fermions the
dimension-5 coupling of the type $(1/M_{P})A_{\mu }\overline{\psi }%
\overleftrightarrow{\partial ^{\mu }}\psi $ satisfying our partial gauge
invariance conjecture could also appear \cite{ckt}. However, for the
coventional SM the minimal Lorentz breaking couplings are proved to be just
the terms presented in the ENSM Lagrangian (\ref{Ltot}).} given in (\ref%
{Ltot}). We showed then that this type of couplings lead, basically through
the deformed dispersion relations of the SM fields, to a new class of
processes being of a distinctive observational interest in high energy
physics and astrophysics some of which have been considered in significant
detail. Such processes leading in themselves to sensitive tests of special
relativity, may also shed some light on a dynamical origin of symmetries
that may only appear, as we argued, if partial rather than exact gauge
invariance holds in the Standard Model.

Though we were mainly focused here on the minimal PGI extension of SM, our
conclusion is likely to largely remain in force for any other extension
provided that they all are determined by the partial gauge invariance
conjecture taken in its general form (\ref{ntt}). It is worth noting,
however, that this conjecture has been solely formulated here for the
hyperchage Abelian symmetry in SM. In this connection, further study of PGI
in a wider context, particularly in conventional Yang-Mills theories and
gravity seems to be extremely interesting.

\section*{Acknowledgments}

One of us (J.L.C.) cordially thanks James Bjorken, Masud Chaichian, Ian
Darius, Colin Froggatt, Roman Jackiw, Oleg Kancheli, Archil Kobakhidze, Rabi
Mohapatra and Holger Nielsen for interesting correspondence, useful
discussions and comments. Financial support from Georgian National Science
Foundation (grants N 07\_462\_4-270 and Presidential grant for young
scientists N 09\_169\_4-270) is gratefully acknowledged by authors.

\end{document}